\documentclass{emulateapj}
\usepackage{graphicx,natbib}
\usepackage{color}

\newcommand{\zphot}{$z_{\rm phot}$}
\newcommand{\zspec}{$z_{\rm spec}$}

\begin{document}

\author{
Hiroaki Oyaizu$^{1,2}$,
Marcos Lima$^{2,3}$,
Carlos E. Cunha$^{1,2}$,
Huan Lin$^{4}$, 
Joshua Frieman$^{1,2,4}$
}

\affil{
${}^{1}$Department of Astronomy and Astrophysics, The University of Chicago, Chicago, IL 60637 \\
${}^{2}$Kavli Institute for Cosmological Physics, The University of Chicago, Chicago, IL 60637 \\
${}^{3}$Department of Physics, The University of Chicago, Chicago, IL 60637 \\
${}^{4}$Center for Particle Astrophysics, Fermi National Accelerator Laboratory, Batavia, IL 60510 \\
}

\title{Photometric redshift error estimators}

\begin{abstract}
Photometric redshift (photo-z) estimates are playing an increasingly important 
role in extragalactic astronomy and cosmology. Crucial to many 
photo-z applications is the accurate quantification of 
photometric redshift errors and their distributions, including  
identification of likely catastrophic failures in photo-z estimates. 
We consider several methods of estimating photo-z errors and propose 
 new training-set based error estimators based on spectroscopic training 
set data. Using data from the Sloan Digital Sky Survey and simulations 
of the Dark Energy Survey as examples, we show that this method provides a 
robust, relatively unbiased estimate of photo-z errors. We show that 
culling objects with large, accurately estimated photo-z errors from a 
sample can reduce the incidence of catastrophic photo-z failures.

\end{abstract}


\section{Introduction}

While spectroscopic redshifts have now been measured
for over one million galaxies, in recent years digital sky
surveys have obtained multi-band imaging for over 
a hundred million galaxies. Deep, wide-area surveys
planned for the next decade will increase the number
of galaxies with multi-band photometry to a few billion.
Over the last decade, substantial 
effort has gone into developing photometric redshift
(photo-z) techniques, which use multi-band photometry
to estimate approximate galaxy redshifts \citep{con95a,bol00,ben00,col04,wad04}. 
For many applications
 in extragalactic astronomy and cosmology, the precision achieved by 
photometric redshifts is sufficient, provided one can accurately 
characterize the uncertainties in the photo-z estimates, i.e., 
the photo-z errors.
A number of recent papers have considered the effects of photo-z
errors on cosmological
probes including baryon acoustic oscillation \citep{zha06}, weak lensing 
tomography \citep{hut06,ma06}, supernovae
\citep{hut04} and galaxy clusters \citep{hut04,lim07}.

A number of methods have been proposed to characterize photometric
redshift errors to date. They 
can be roughly divided into two categories:
methods based on estimating statistical errors in template fitting, e.g., 
the $\chi^{2}$ method and its Bayesian
counterparts \citep{bol00,ben00};  
and methods that explicitly propagate errors in the input
parameters, typically magnitudes or colors, through the photo-z 
estimator \citep[e.g.,][]{bru99b,hsi05,col04}. 

The error in a photometric redshift estimate \zphot~ is simply the difference 
between the photo-z estimate and the true (hereafter, spectroscopic) redshift,
$\Delta z =$\zphot $-$\zspec. In practice, the errors for the vast 
majority of objects in a deep photometric sample are unknown, since the 
spectroscopic redshifts are not measured. Our goal is to devise an 
estimator of $\Delta z$ that has desirable statistical properties, e.g., minimum 
bias and variance, based on whatever information is at hand. Given a photo-z 
estimate, an error estimator should give the range of redshifts over which the 
true redshift will be found at some confidence level. 

In most cases, spectroscopic redshifts are available for a small 
subset of the photometric sample. Such spectroscopic samples are 
often used as training sets for empirical or 
machine-based learning photo-z estimators. 
In this paper, we develop methods of 
photo-z {\it error} estimation that are based on the use of spectroscopic 
training sets to accurately characterize the error distribution.
We show that training-set based error estimators
 outperform other commonly used methods when a representative 
training set is available and that they are competitive even when 
the training set is not fully representative of the photometric sample.
In cases where the magnitude errors are not well determined,
we show that the relative advantages of the new training-set
based methods are further increased.

This paper is organized as follows.
In \S \ref{section:test}, we describe the data sets that we use
in this work.
In \S \ref{section:train}, we introduce the training-set based error
estimators and their implementations, as well as their advantages and
disadvantages.
For comparison, we review the traditional error estimators in \S 
\ref{section:comparison} and highlight the key differences between
them and the training-set based error estimators.
We show in \S \ref{section:culling} that the over-all photo-z 
scatter and outlier fraction can be
significantly improved by culling objects with high estimated photo-z 
errors, possibly leading to improved results in analyses that rely on
photo-z's.
Finally, we offer concluding remarks in \S \ref{section:conclusion}.

\section{Test Methods and Data}\label{section:test}
\begin{figure}
  \begin{center}
    \begin{minipage}[t]{85mm}
      \resizebox{85mm}{!}{\includegraphics[angle=0]{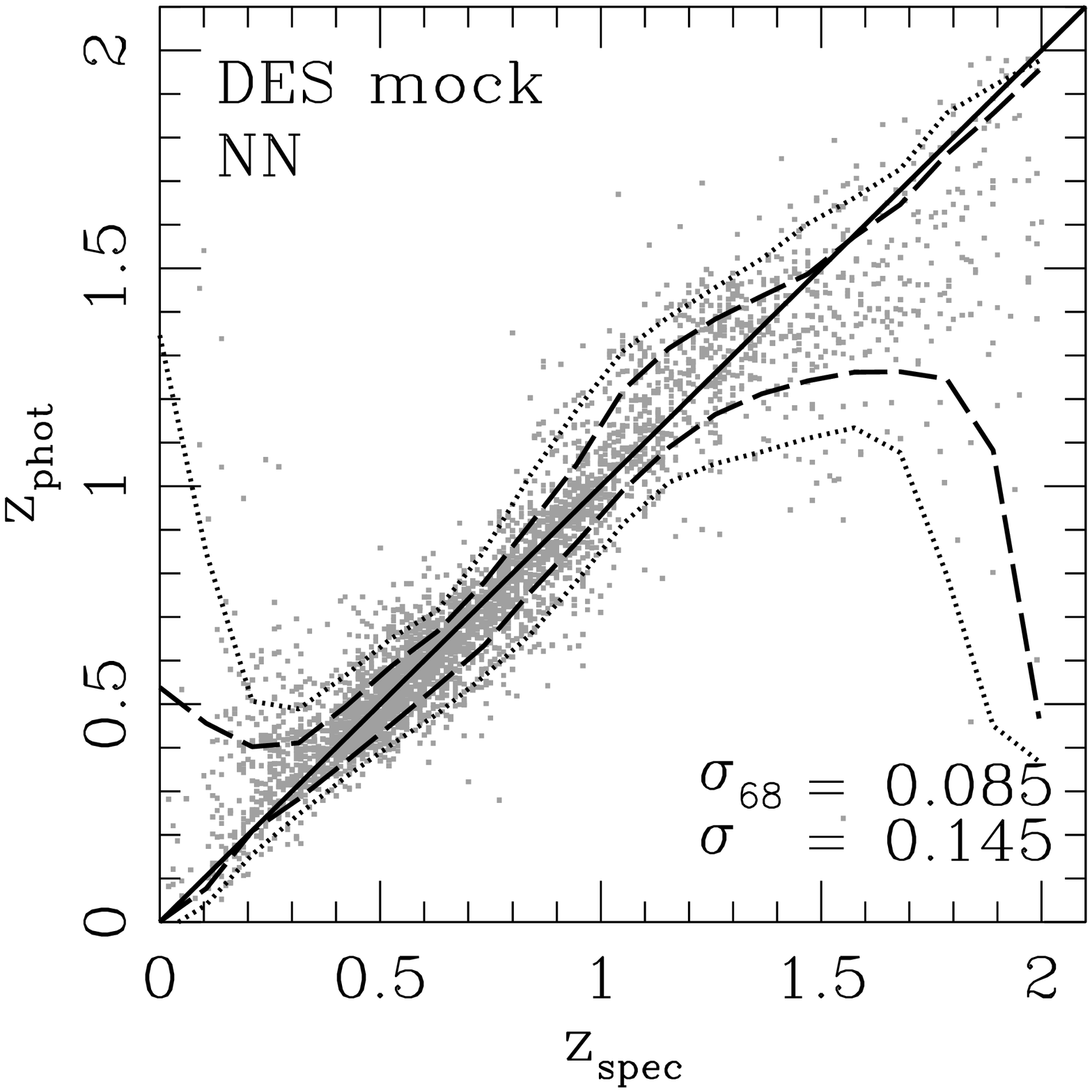}}
    \end{minipage}
    \begin{minipage}[t]{85mm}
      \resizebox{85mm}{!}{\includegraphics[angle=0]{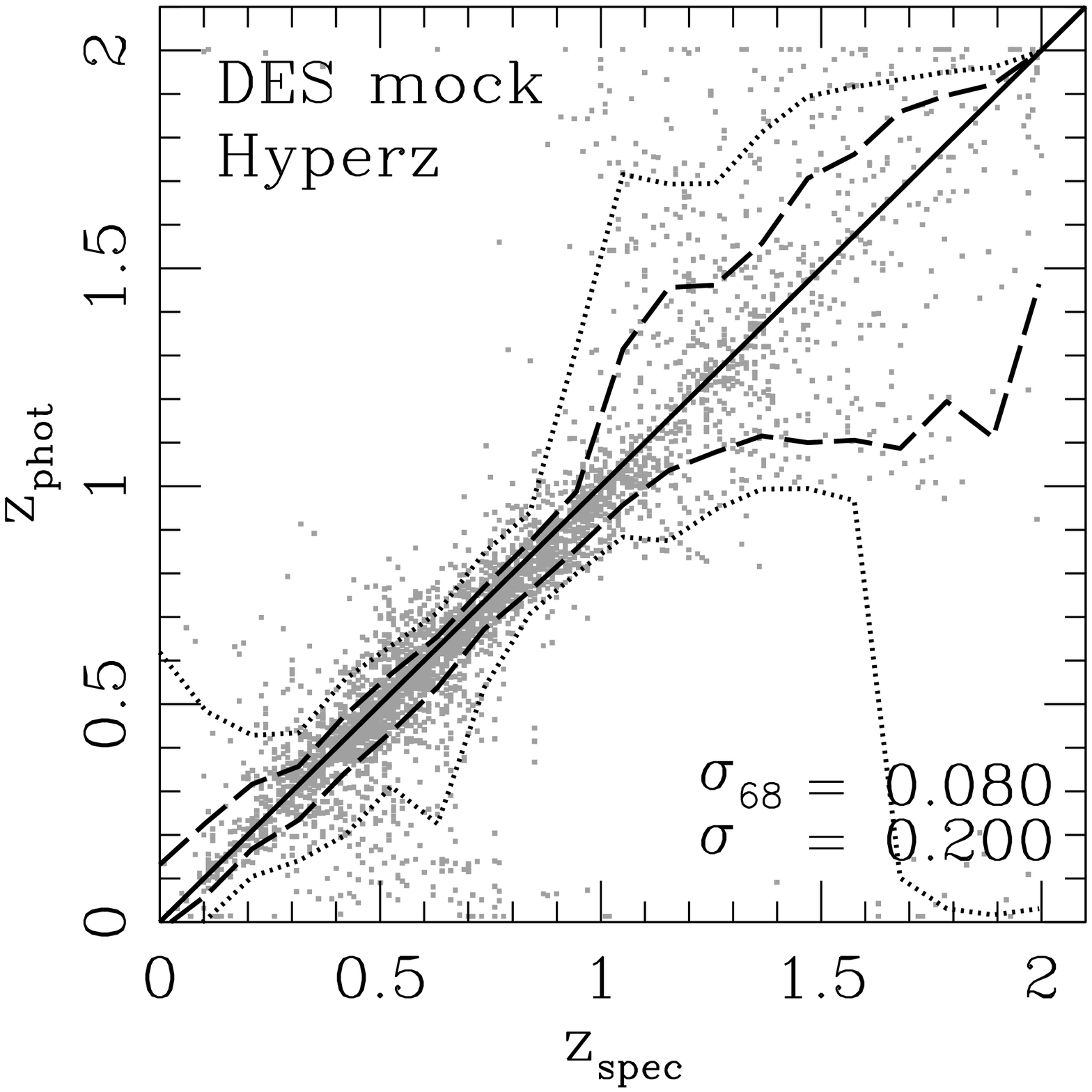}}
    \end{minipage}
    \caption{Photometric versus spectroscopic redshift for the DES mock catalog
      photometric set, 
      calculated using the Neural Network ({\it top panel}) 
      and Hyperz ({\it bottom panel}) methods.  The dashed and dotted curves enclose 
      68\% and 95\% of the points in each \zspec~ bin. In the lower right of 
      each panel, $\sigma$ is the {\it rms} photo-z scatter averaged over all 
      $N$ objects in the photometric set 
      , $\sigma^2=(1/N)\Sigma_{i=1}^{N} 
      (\Delta z^i)^2$, and $\sigma_{68}$ is the range containing 68\% of the 
      validation set objects in the distribution of $\Delta z$.
      The Hyperz photo-z's for the DES mock catalog are calculated with 
      $z_{\rm max}$ set to 2.
    }
    \label{plot:zpzsDES}
  \end{center}
\end{figure} 

\begin{figure}
  \begin{center}
    \begin{minipage}[t]{85mm}
      \resizebox{85mm}{!}{\includegraphics[angle=0]{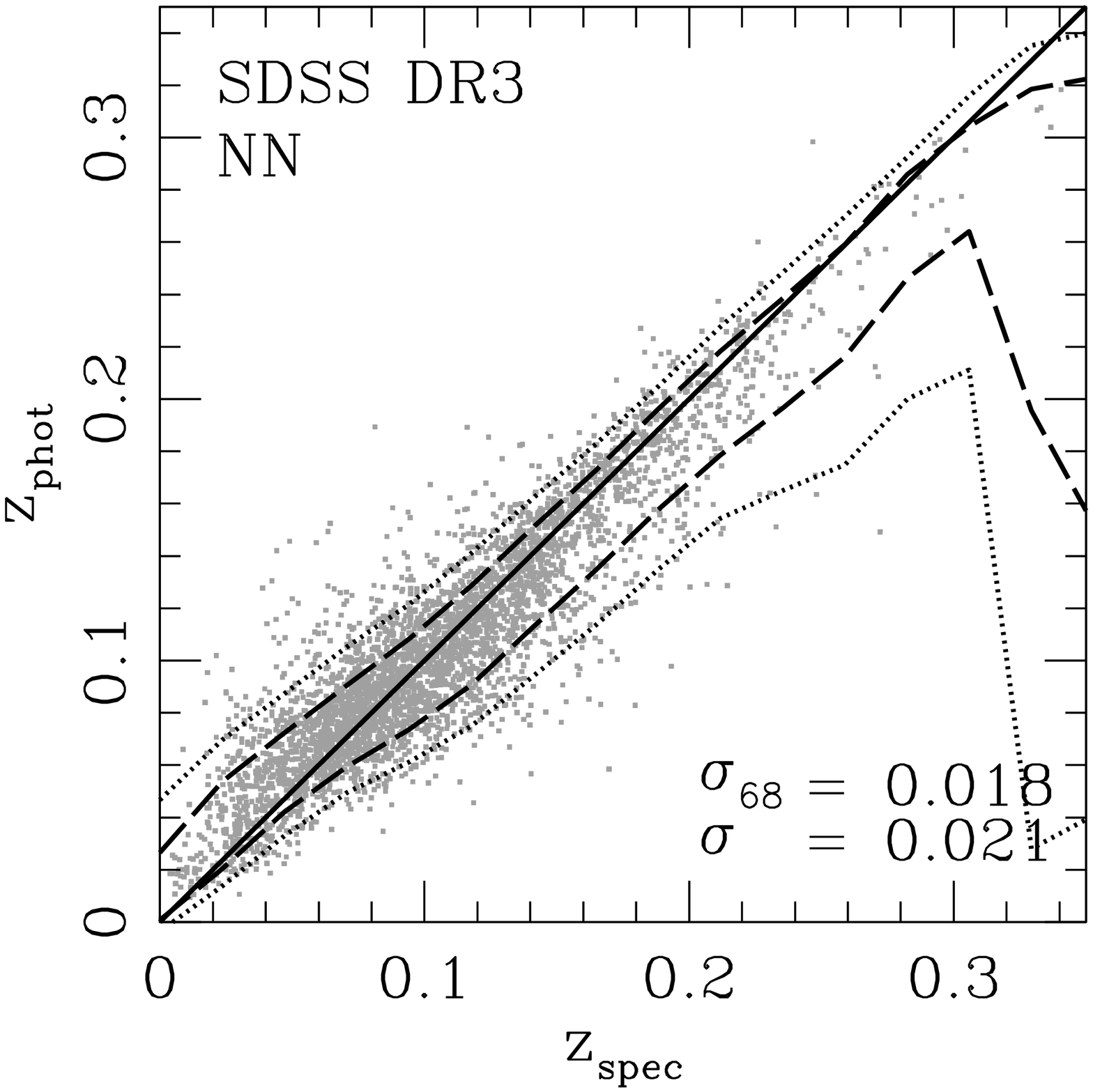}}
    \end{minipage}
    \begin{minipage}[t]{85mm}
      \resizebox{85mm}{!}{\includegraphics[angle=0]{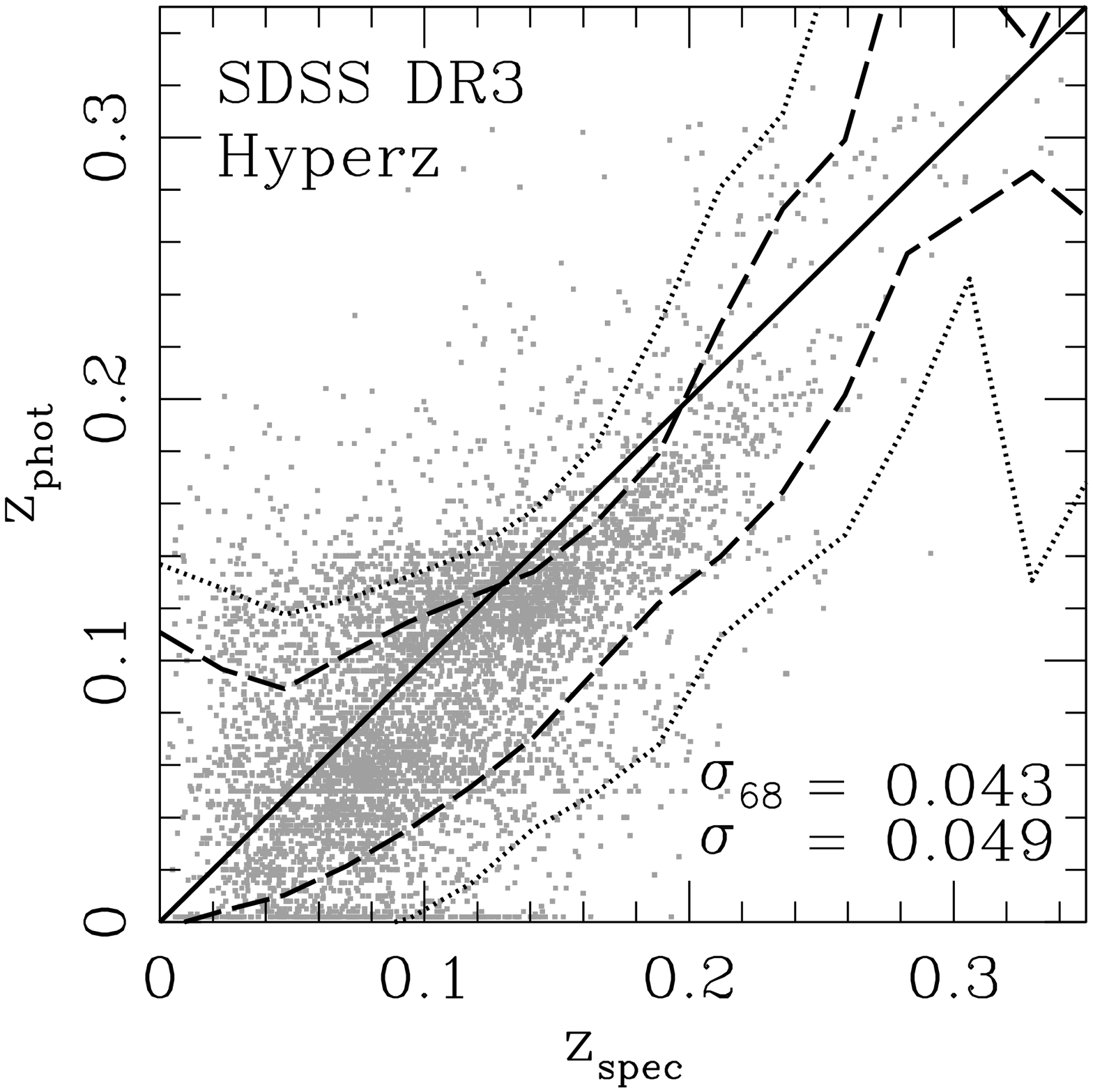}}
    \end{minipage}
    \caption{Photo-z versus spectroscopic redshift for the SDSS DR3 photometric set 
      calculated using the Neural Network ({\it top panel})
      and Hyperz ({\it bottom panel}). 
      The Hyperz photo-z's for the SDSS catalog are calculated with 
      $z_{\rm max}$ set to 0.4.
    }
    \label{plot:zpzsSDSS}
  \end{center}
\end{figure}

In order to fairly compare the qualities of various photometric redshift
error estimators, we have compiled two galaxy photometric catalogs. 
Each catalog consists of spectroscopic redshifts, magnitudes in several 
chosen filter passbands, and magnitude errors.

The first catalog is a simulated data set created to resemble observations of 
the proposed Dark Energy Survey (DES) \citep{des}. 
The DES is a 5000 square degree survey in 5 optical passbands ($grizY$) with a 
magnitude limit of $i \approx 24$, to be carried out using a new camera on 
the CTIO 4-meter telescope.
The goal of the survey is to measure the equation
 of state of dark energy using several techniques: clusters, 
weak lensing, angular galaxy clustering (baryon acoustic oscillations), 
and supernovae. Since DES will observe $\sim 300$ million galaxies, 
the redshifts must be obtained using photometric methods.
The DES optical survey will be complemented in the near-infrared by the VISTA Hemisphere Survey, 
an ESO Public Survey on the VISTA 4-meter telescope that will cover the survey area in 
$J, H$, and $Ks$. While the color information provided by $grizYJHKs$ photometry leads 
to improved photo-z estimates compared to optical-only imaging, for simplicity and purposes of 
illustration the mock catalog we use here contains only $griz$ magnitudes.

The simulated DES catalog contains 200,000 galaxies with $z < 2$ and 
with $20 < i < 24$.
The magnitude and redshift distributions are derived from the galaxy luminosity 
function measurements of \cite{lin99} and \cite{pol03}, while the galaxy SED type 
distribution is obtained from measurements of the 
HDF-N/GOODS field \citep{cap04,wir04,cow04}. 
The galaxy colors are generated using the four \citet{col80} templates--E, Sbc, Scd, 
Im--extended to the UV and NIR using synthetic templates from \citet{bru93}. 
To improve the sampling and coverage of color space, we create additional templates 
by interpolating between adjacent templates or by extrapolating from the E and Im templates.

The second test catalog we use is based on the Sloan Digital Sky Survey (SDSS) 
Data Release 3 \citep{aba03}. Although this catalog has been superceded 
by later data releases \citep{ade07}, for which we have published 
a photo-z catalog \citep{oya08}, 
it nevertheless provides a useful testbed for 
studies of photo-z errors.
This SDSS catalog contains spectroscopic redshifts and magnitudes in 
$ugriz$ passbands for $292,964$ galaxies from the main spectroscopic 
sample, which is flux-limited to $r < 17.77$.
Because this sample is confined to low redshift, $z \lesssim 0.3$, 
most of the strong features of galaxy spectra targeted by photometric 
redshift estimators fall within the wavelength range covered by the filters.
A notable exception is the Lyman alpha emitters at $z > 2.5$.
However, the fraction of these high redshift objects in our sample is too
small to have measurable effects on our results.

We calculate photometric redshifts for these catalogs using two
methods, a Neural Network (NN)  
method and the $\chi^{2}$ based spectral template
fitting package Hyperz \citep{bol00}.
The NN technique is a training-set method based on fitting a parametrized
function, represented by a feed-forward multilayer perceptron (FFMP) 
neural network, to the redshift-magnitude relation embodied in a 
spectroscopic training set. 
The implementation is the same as the one described in \citet{oya08} for 
the SDSS DR6 photo-z catalog,
except that the network configurations are different: 
here we use a 4:15:15:15:1 network
for the DES catalog and a 5:15:15:15:1 network for the SDSS catalog.
Figures \ref{plot:zpzsDES} and \ref{plot:zpzsSDSS} show the resulting
photometric redshifts plotted against spectroscopic redshifts for all
catalogs used in this study.

We split 
the DES and SDSS catalogs into three independent
catalogs each, labeled training, validation, and photometric sets.
The sizes of these sets are 50,000, 50,000, and 100,000 for the DES and
100,000, 92,964, and 100,000 for the SDSS.
Except where noted below (\S \ref{extrap}), these subsets are drawn at random from the  
photometric samples, i.e., they are each statistically representative of the 
full samples. 
When the photo-z's are determined using the NN training-set method, we use
the training and validation sets to determine the mapping from magnitudes
to redshifts and magnitudes to redshift errors. 
The resulting mapping is then applied to the photometric set for comparison
of the training-set error estimator against other error estimation methods. 
Splitting the catalogs ensures that the training-set error methods are not 
given unfair advantage with respect to the other error estimators. When we 
estimate photo-z's and photo-z errors using template methods, we apply 
the methods directly to the photometric set.

\section{Error estimates using training sets}\label{section:train}

Training set based photo-z estimators \citep[e.g.,][]{con95a,csa03,col04} 
use a spectroscopic training set, typically a subset of the photometric 
sample, to derive a functional relation between 
redshift and photometric observables (e.g., magnitudes) which is then 
applied to the photometric sample of interest. In the same spirit, 
we can also use a training set to derive an estimate of the photo-z {\it error}, 
that is, a relation between photo-z error and some photometric observables. 
Note that the error estimator does not need to make use of the same 
observables as the photo-z estimator. In fact, we stress that the empirical photo-z 
error estimators are independent of the method used to estimate 
photometric redshifts themselves: training-set based error estimators can 
be applied to either empirical (training set) or template-based photo-z 
estimates. The assumption underlying the training-set based error estimator 
is that there is a functional relationship between some set of photometric 
observables and photo-z error and that this relationship for the 
training set data is reasonably representative of the relationship for 
the photometric sample as a whole.

In the following subsections, we describe and test two basic techniques 
that use a spectroscopic training set to estimate photo-z errors. Both techniques are 
based on the simple observation that objects 
with similar magnitudes in a photometric survey tend to have 
similar photometric errors, and such magnitude errors are typically 
the largest contributors to photometric redshift error. Therefore, 
objects with similar multi-band magnitudes will tend to have similar 
photo-z errors. Moreover, such neighbors in magnitude space, having 
similar colors, usually (but not always) correspond to galaxies with similar SEDs.  
Photo-z errors depend strongly on SED type, since the quality of photo-z estimates 
is related to the presence of strong and broad spectral features.
We can therefore group objects in a spectroscopic training set according 
to their magnitudes and determine the photo-z error as a function of 
the magnitudes using the training set.
For each object in the photometric set, we then find the objects in 
the training set that are near it in magnitude-space and associate some weighted mean of 
the measured errors for these training-set neighbors to it.
The two methods introduced below differ in the method of grouping the 
galaxies.

\begin{figure*}
  \begin{center}
    \begin{tabular}{cccc}
      \resizebox{40mm}{!}{\includegraphics[angle=0]{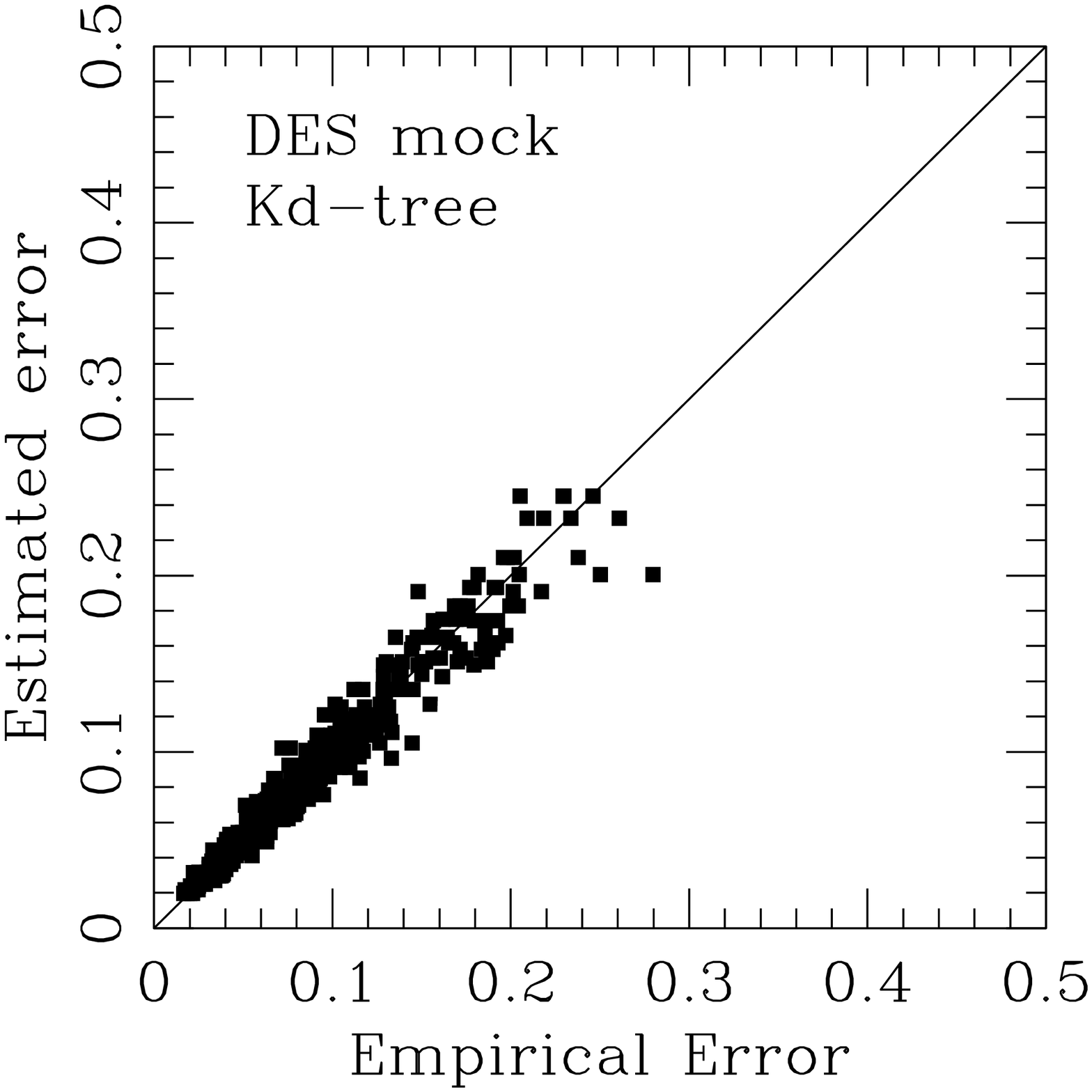}} &
      \resizebox{40mm}{!}{\includegraphics[angle=0]{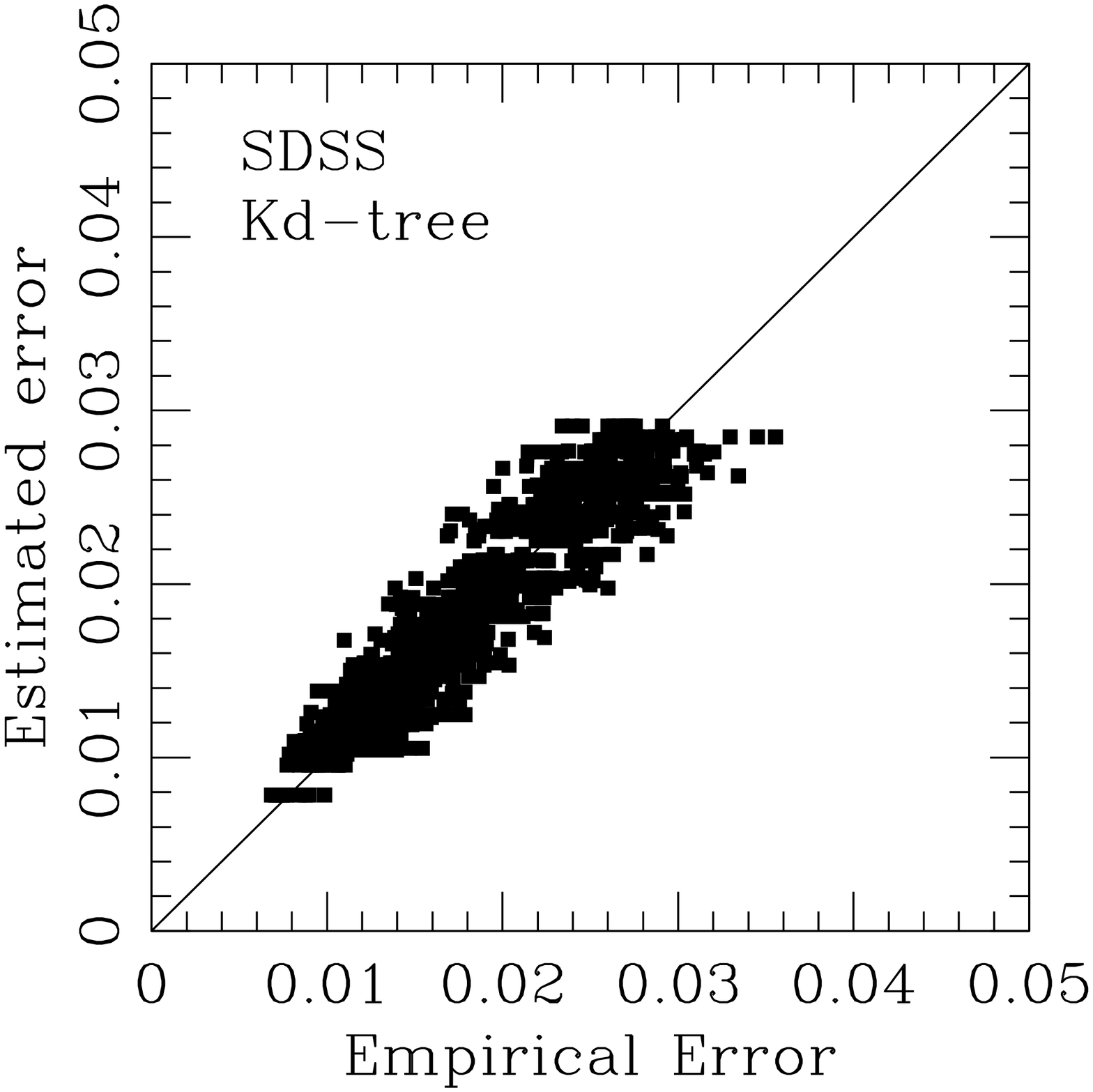}} &
      \resizebox{40mm}{!}{\includegraphics[angle=0]{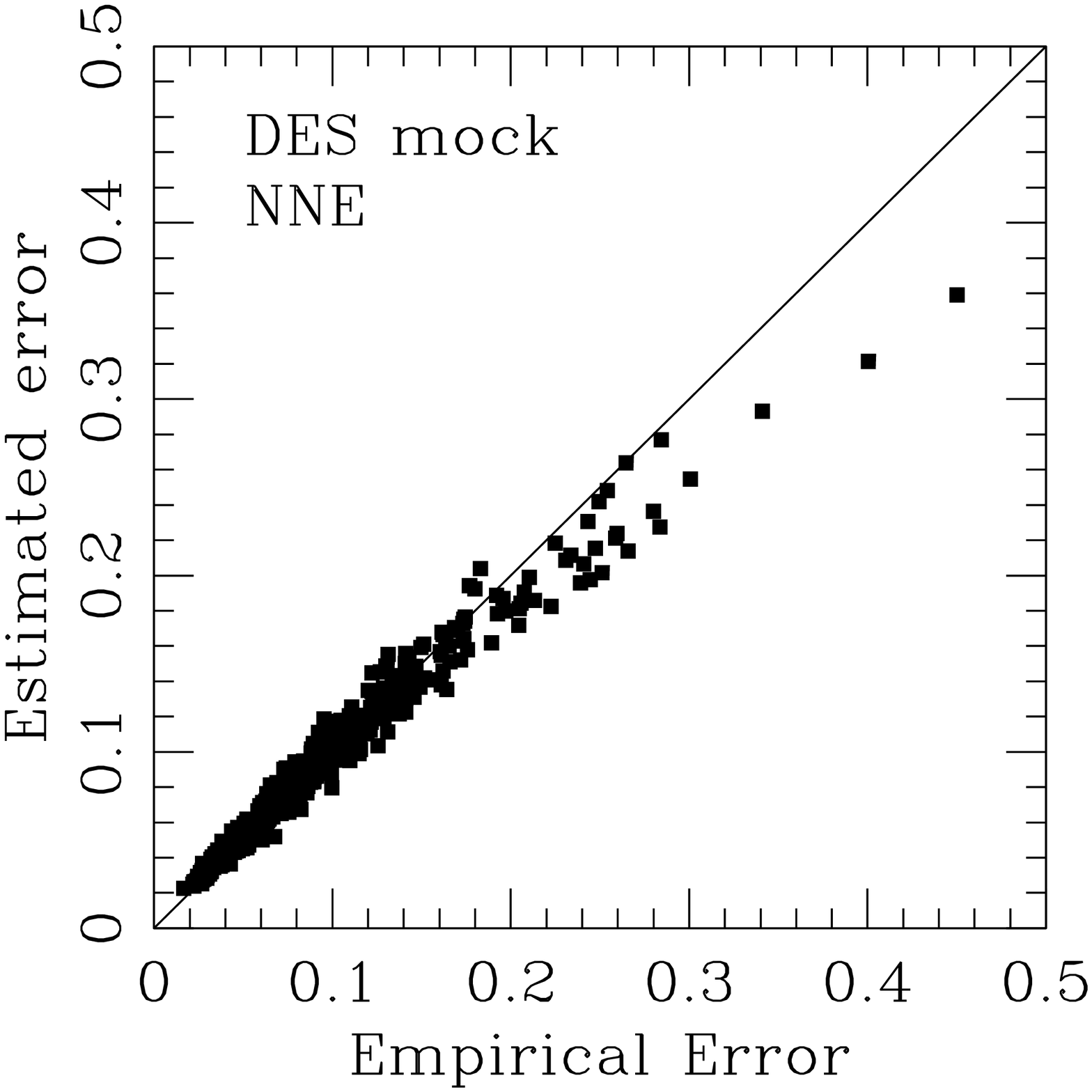}} &
      \resizebox{40mm}{!}{\includegraphics[angle=0]{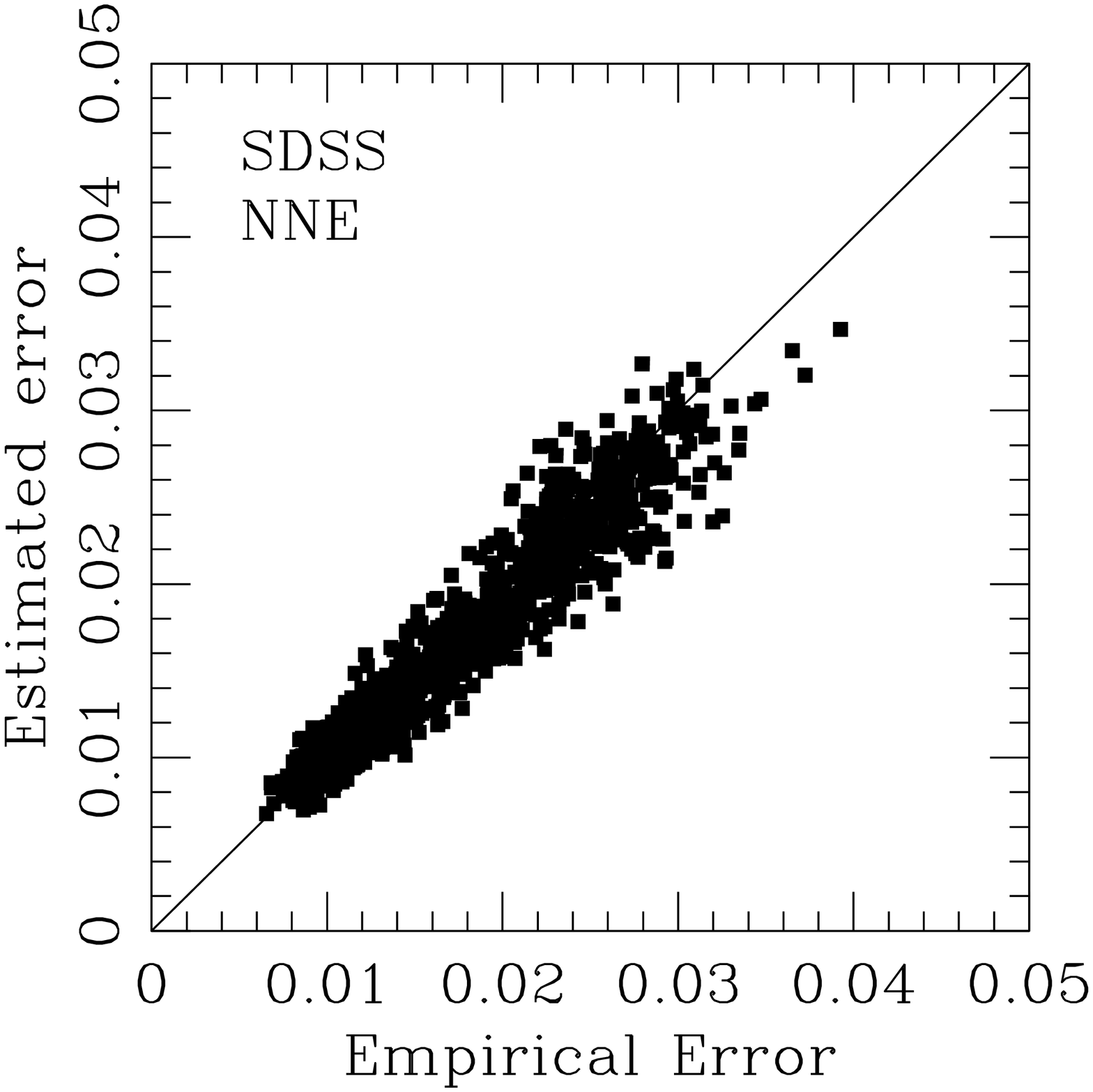}} \\
      \resizebox{40mm}{!}{\includegraphics[angle=0]{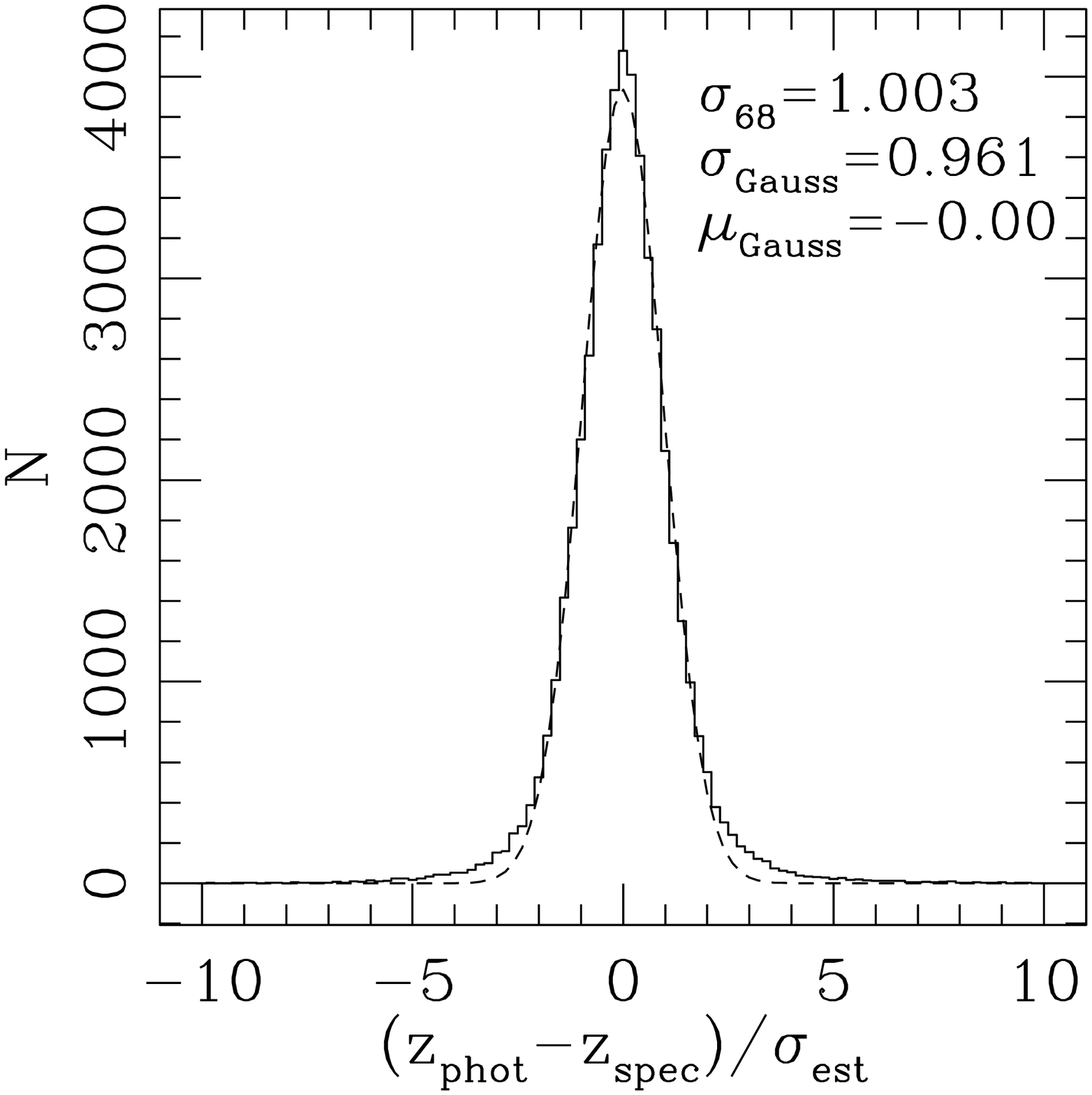}} &
      \resizebox{40mm}{!}{\includegraphics[angle=0]{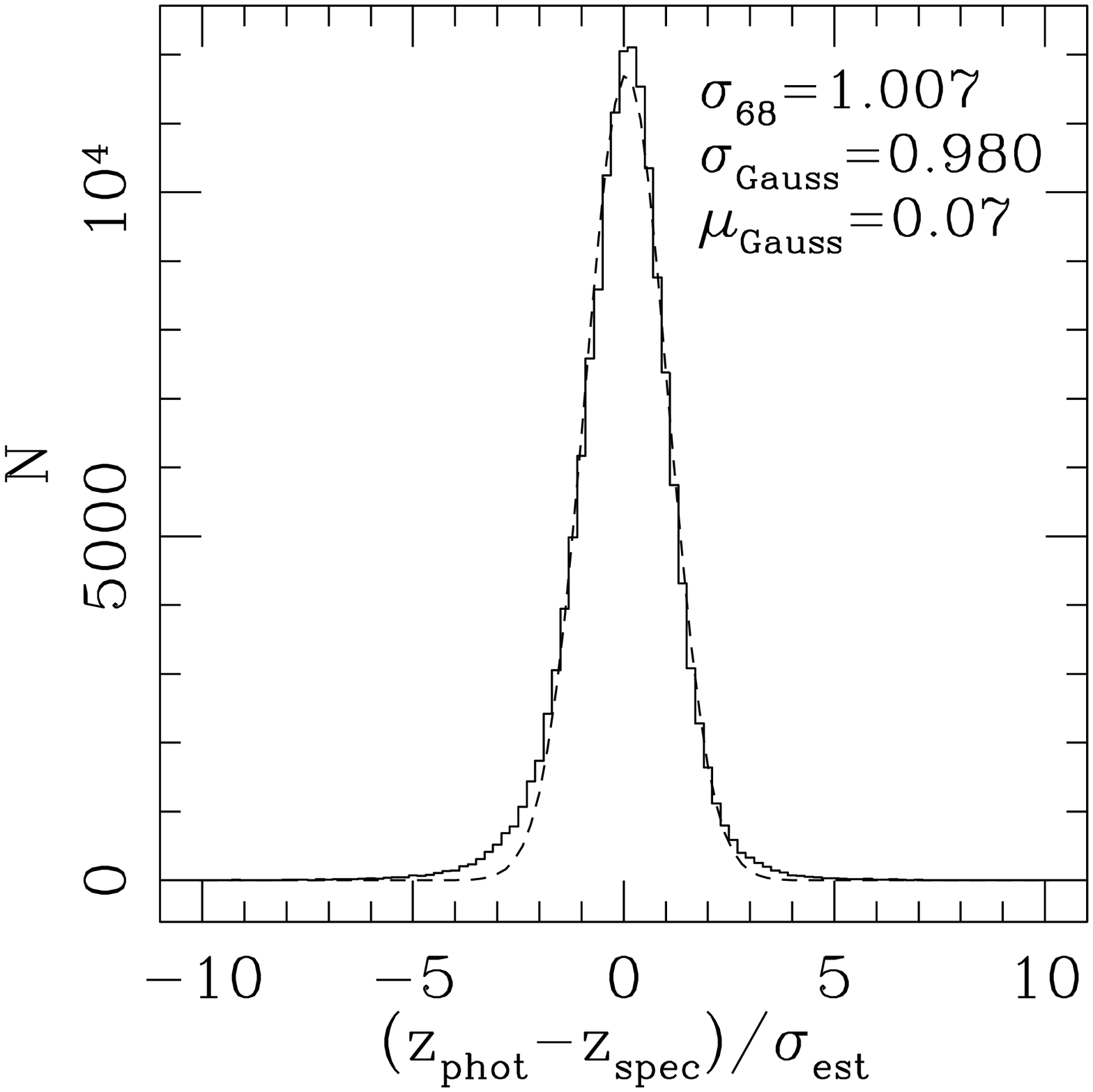}} &
      \resizebox{40mm}{!}{\includegraphics[angle=0]{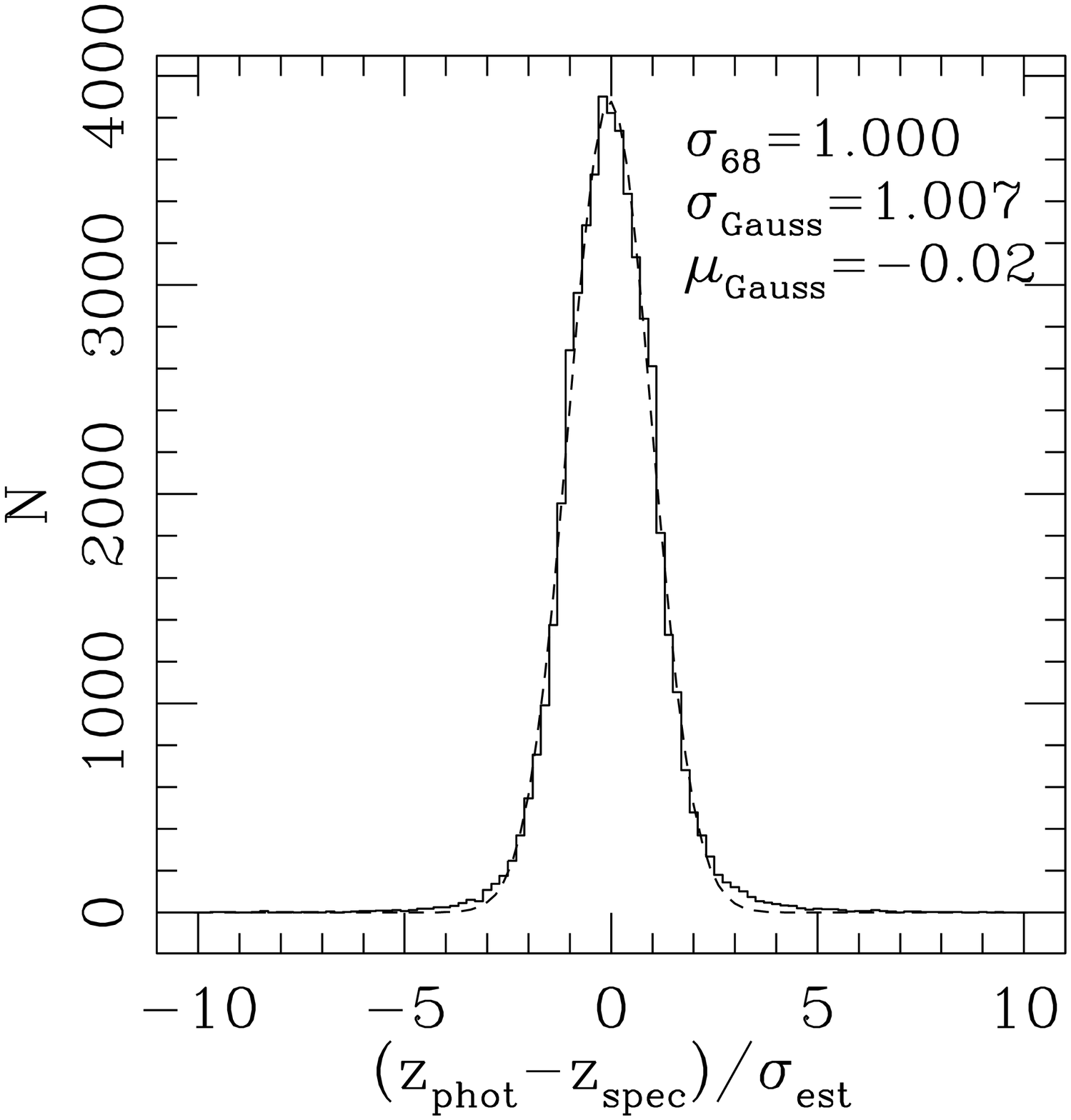}} &
      \resizebox{40mm}{!}{\includegraphics[angle=0]{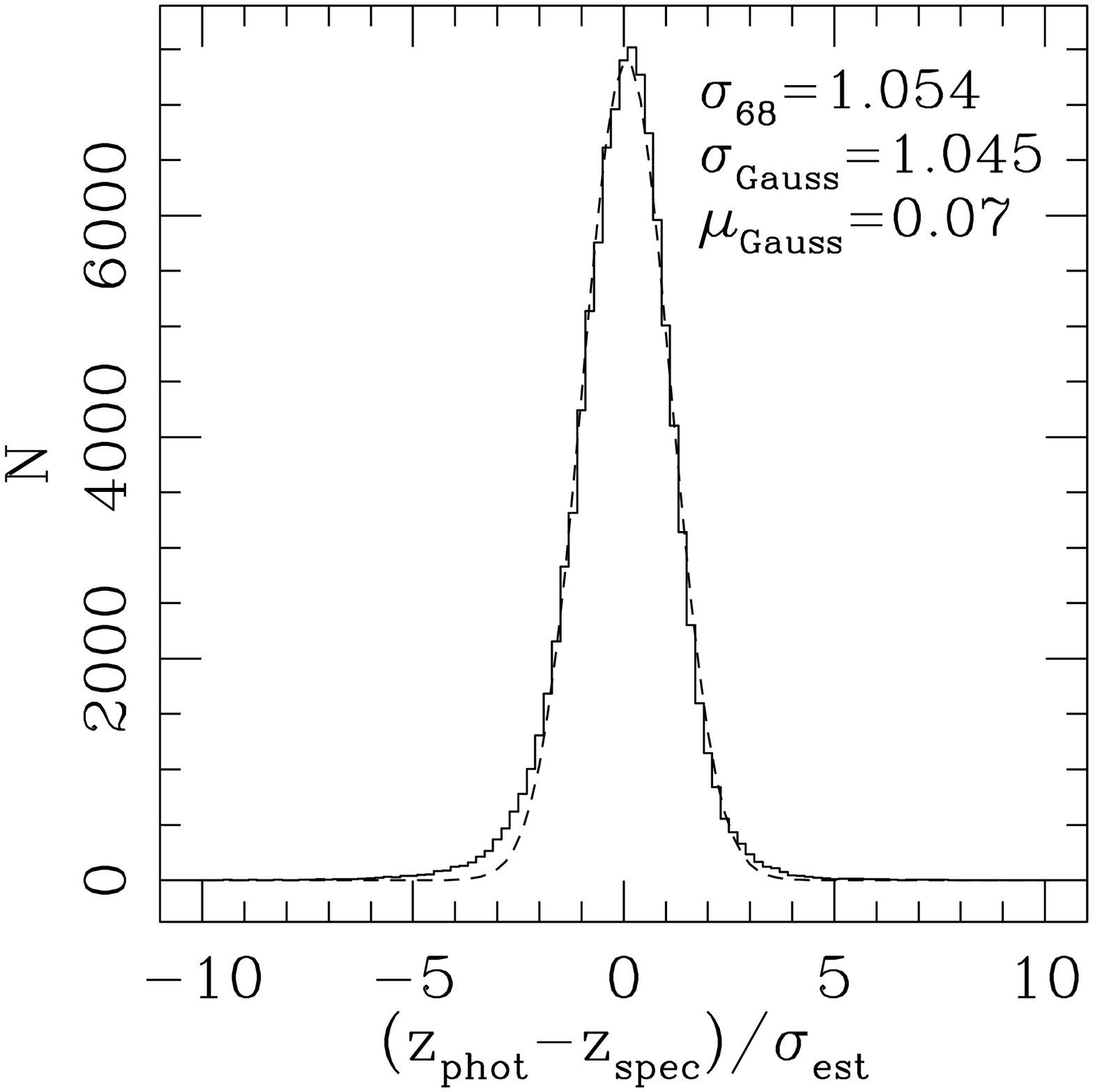}} \\
    \end{tabular}
  \end{center}
  \caption{{\it Upper panels:} Estimated error vs empirical error for 
a) DES photometric set using Kd-tree error estimate, b) SDSS using Kd-tree, c) DES 
using Nearest Neighbor error (NNE)  
estimate, and d) SDSS using NNE. In all four cases, the neural network method was used to 
estimate the photo-z's. 
{\it Lower panels:} Corresponding distributions of $(z_{\rm phot}-z_{\rm spec})/\sigma_{\rm est}$, 
where $\sigma_{\rm est}$ is the photo-z error estimate for each galaxy. Solid histograms 
show the distributions, dashed curves are Gaussian fits to the distributions. 
\label{train1}} 
\end{figure*}

\subsection{Kd-tree Error Estimator}

The first photo-z error method we consider uses a Kd-tree algorithm to 
bin training-set objects in magnitude space.
A Kd-tree (short for K-dimensional tree) is a general data organization and
classification algorithm that is suited for efficiently partitioning data 
points in multi-dimensional parameter spaces.
In our implementation, 
the training set is partitioned into two bins at the median value of the first photometric parameter 
(which we choose to be $u$ mag for SDSS, $g$ for DES).
For each bin, the objects within the bin are further partitioned
at the median of the second parameter (here $g$ for SDSS, $r$ for DES), 
resulting in $2^2 = 4$ bins. This process is continued for the photometric parameters 
of interest (here the 5 magnitudes for SDSS and 4 for DES). We then return  
to the first parameter, partition each bin at the median of the first parameter 
for that bin, cycle again through the parameters, and continue subdividing until
the number of objects in a bin becomes sufficiently small.
Once the partitioning is completed, we calculate the 68\% width of the error 
distribution centered about $z_{\rm phot}-z_{\rm spec} = 0$ for each bin 
and declare that to be the photo-z error estimate for objects in the 
{\it photometric} sample that fall within that bin.

Because the Kd-tree bins are always partitioned at the median value of 
the object distribution in some parameter, the number of training-set 
objects per bin, $N_b$, is nearly constant from bin to bin. 
This constancy ensures a nearly uniform shot-noise uncertainty 
($\propto 1/\sqrt{N_b}$) on the estimates of the photo-z errors. 
While this statistical 
uncertainty is minimized by having many objects per bin,  
large bins are ``non-local'' in multi-magnitude space, and the training-set
based error estimator is predicated on the locality assumption that similar 
magnitudes imply similar errors. Therefore, the optimal bin size 
should be as small as possible (or smaller than 
the scale over which the error distribution changes appreciably)  
but large enough that the 
shot-noise error is not large compared to the error induced by
non-locality of the bin. 
For the training set samples we consider here, we find that 
$N_b \simeq 100$ objects per bin is nearly optimal.
The size of the training set can also change the locality of the 
nearest $N_b$ neighbors, and in general, the required locality depends
on the first derivative of the redshift-magnitude relationship.
Because such relationships are dependent on numerous factors, such as 
filter choice, selection function, and magnitude errors, we cannot
provide a general requirement for the training set size.
We note, however, that in both DES mock and SDSS catalogs, we find
virtually no improvement in error estimator quality when the training set
size is larger than 20,000 galaxies.

Figures \ref{train1}a and \ref{train1}b show the results of  
applying the Kd-tree error estimator to the DES and SDSS photometric sets. In these 
cases, the neural network (NN) method was used for the photo-z estimates.  
The photo-z errors are estimated using a Kd-tree with 512 bins 
for the DES catalog and 1024 bins 
for the SDSS catalog, corresponding to $N_b \simeq 97$ training-set objects 
per bin in each case. The top panels of 
Figure \ref{train1} show the photo-z error estimates 
vs.\ the measured or ``empirical'' errors.
In order to compute the empirical error, we first sort the galaxies
according to their estimated error.
Next, we bin the galaxies into bins of 100 objects starting from the 
galaxy with the smallest estimated error, and call the average estimated
error of the galaxies within a bin the ``estimated error'' of the bin,
which is plotted on the vertical axis of Figure \ref{train1}.
Finally, we compute the 68\% width of the $|z_{\rm phot}-z_{\rm spec}|/
\sigma_{\rm est}$ distrubtion of each bin, and call it the
``empirical error'' of the bin.
The assumption here is that if the error estimator is working properly,
 those objects with similar estimated error should follow similar underlying
 error distributions, and the underlying distribution should have a width
 that is well-approximated by the estimated error.
As the figure shows, the estimated Kd-tree error correlates well 
with the true error, with almost no apparent bias and relatively 
small scatter.

The solid histograms in the lower panels of 
Figure \ref{train1} show the corresponding distributions of 
$(z_{\rm phot}-z_{\rm spec})/\sigma_{\rm est}$, 
where $\sigma_{\rm est}$ is the Kd-tree error estimate. 
The dashed curves in these panels show 
Gaussian fits to the error distributions; we also 
indicate the best-fit Gaussian
means ($\mu_{\rm Gauss}$) and standard deviations 
($\sigma_{\rm Gauss}$) as well as the $\sigma_{68}$ widths (about zero) of the
distributions (not the fits). 
The fits give equal weight to each bin of the distributions and 
ignore objects for which $\sigma_{\rm est}=0$.
There is no {\it a priori} 
reason for these error distributions to be Gaussian.
Nevertheless, for the Kd-tree error estimator, the error distributions are very close to 
Gaussians, except for small tails seen for both the DES and
SDSS catalogs.
The tails are signatures of catastrophic photo-z failures: due to 
photometric errors, an intrinsically underluminous, 
red galaxy at low 
redshift, for example, may ``scatter into'' a bin mostly populated (in the 
training set) by intrinsically luminous, blue galaxies at much higher redshift. 
In such degenerate cases, the photo-z error is large, and 
the Kd-tree error underestimates the true error: in this example, the  
Kd-tree error assigned to the red galaxy interloper would be dominated by the small errors of the 
blue galaxies in that bin. With a sufficiently large training set, one could hope to 
identify such problematic bins in magnitude space, since the photo-z error distributions in the 
training set for those bins would show anomalous tails.

A disadvantage of the Kd-tree method is the fact that the estimated error 
is discrete.
There can only be as many different error estimates as there are Kd-tree bins, 
and this limits the resolution of the estimated photo-z errors, 
especially for objects with large photo-z errors as seen by the lack of
high Kd-tree estimated errors in Figure \ref{train1}. 
This problem can in principle be alleviated by using more Kd-tree bins.
However, as noted above, for fixed training set size, 
the number of bins is limited by the requirement that each bin should 
contain enough training-set objects to determine the error with small 
shot-noise uncertainty.

\subsection{Nearest Neighbor Error estimator}\label{nnesection}

While the Kd-tree error estimator was seen to have good statistical 
properties, we have found that a Nearest Neighbor Error (NNE) estimator 
performs even better. 
Note that the NNE has in principle nothing to do with Neural Networks (NN),
and the readers should be careful not to confuse the similar acronyms.
In this method, for each object in the photometric set, we estimate the photo-z 
error by using the 68\% spread of the error distribution of its $N_{\rm nei}$ nearest 
neighbors in the training set. Here, nearness in magnitude space is defined 
using the Euclidean metric: 
given two objects with two 
sets of measured magnitudes ${\textbf m_1}$ and ${\textbf m_2}$, 
we define the distance between them by   
\begin{equation}
D^2 = |{\bf m_1} - {\bf m_2}|^2 = \sum_{\mu=1}^{N_m} (m_1^{\mu} - m_2^{\mu})^2 ~,
\end{equation}
\noindent where $N_m$ denotes the number of magnitudes (different passbands) 
measured for each object. In contrast to the Kd-tree method, in NNE 
each object in the photometric set defines its own ``bin.''

The choice of the number of nearest neighbors ($N_{\rm nei}$) to use is analogous to 
the choice of the number of bins in the Kd-tree error estimate.
We prefer to keep the number of neighbors constant for all objects in the 
photometric set, since the shot noise of the resulting error
estimate is then fixed. As with the Kd-tree method, one should choose $N_{\rm nei}$ 
large enough to keep the shot noise of the estimate under control but small 
enough so that the error estimate remains relatively local in magnitude space. 
For the samples we have tested in this analysis, we again find that 
$N_{nei} \simeq 100$ training-set neighbors is nearly optimal.

The upper panels of 
Figures \ref{train1}c and \ref{train1}d show the results of applying the 
NNE estimation method to the DES and SDSS catalogs.
The discreteness that was a concern for the Kd-tree error estimate is not present 
in the NNE method. Moreover, 
the NNE error displays tighter correlation with the empirical error, 
because a nearest-neighbor bin for a photometric object is almost always 
more local in magnitude space than a Kd-tree bin for the same object. The lower panels 
of the same Figures show that 
the error distributions are reasonably well fit by Gaussians, with 
widths that are within 5\% of the expected width $\sigma_{\rm Gauss} = 1$.
Non-Gaussian tails similar to those seen in the Kd-tree error distributions
are also present in the NNE error distributions, for the same reasons.

\begin{figure}
  \begin{center}
    \begin{tabular}{cc}
      \resizebox{40mm}{!}{\includegraphics[angle=0]{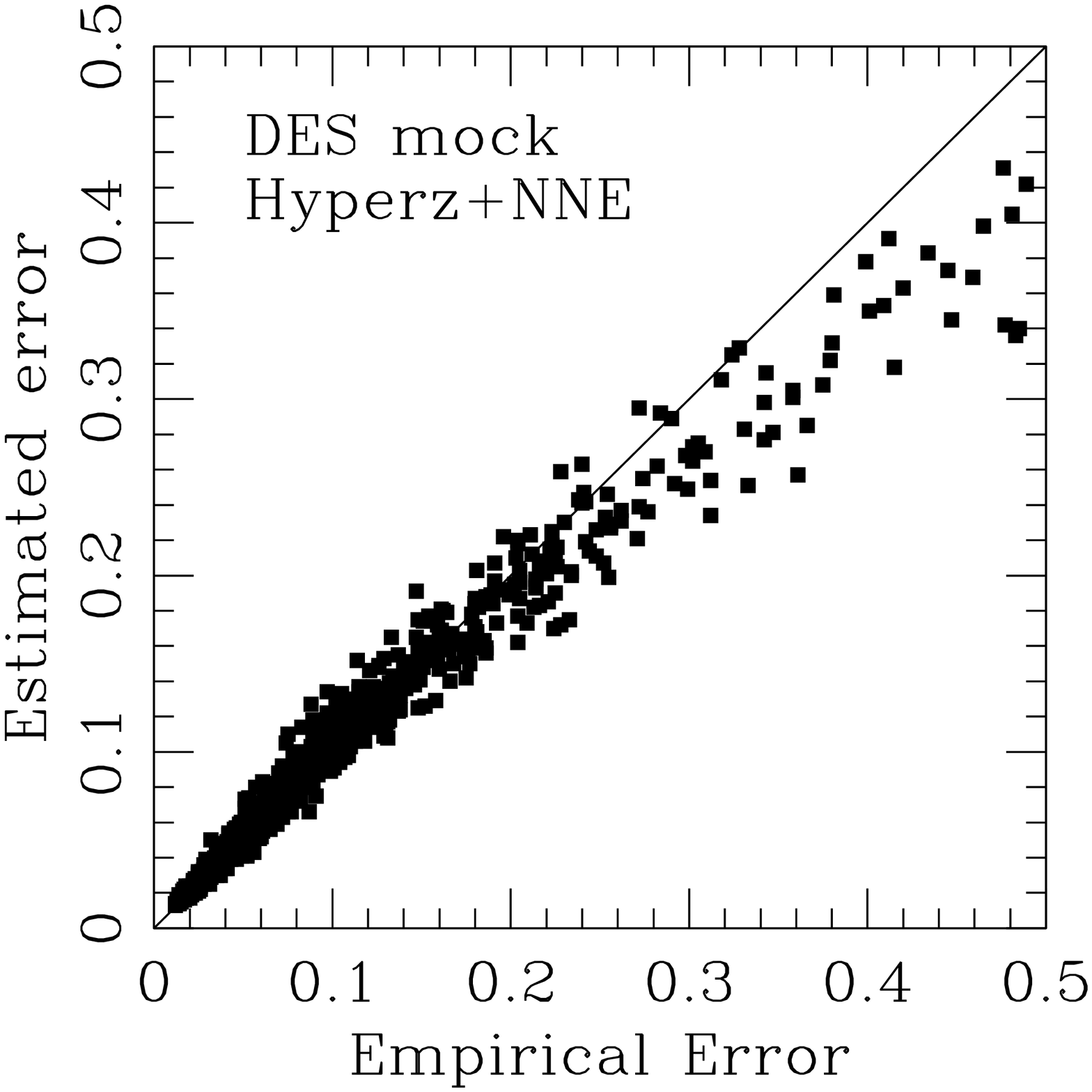}} &
      \resizebox{40mm}{!}{\includegraphics[angle=0]{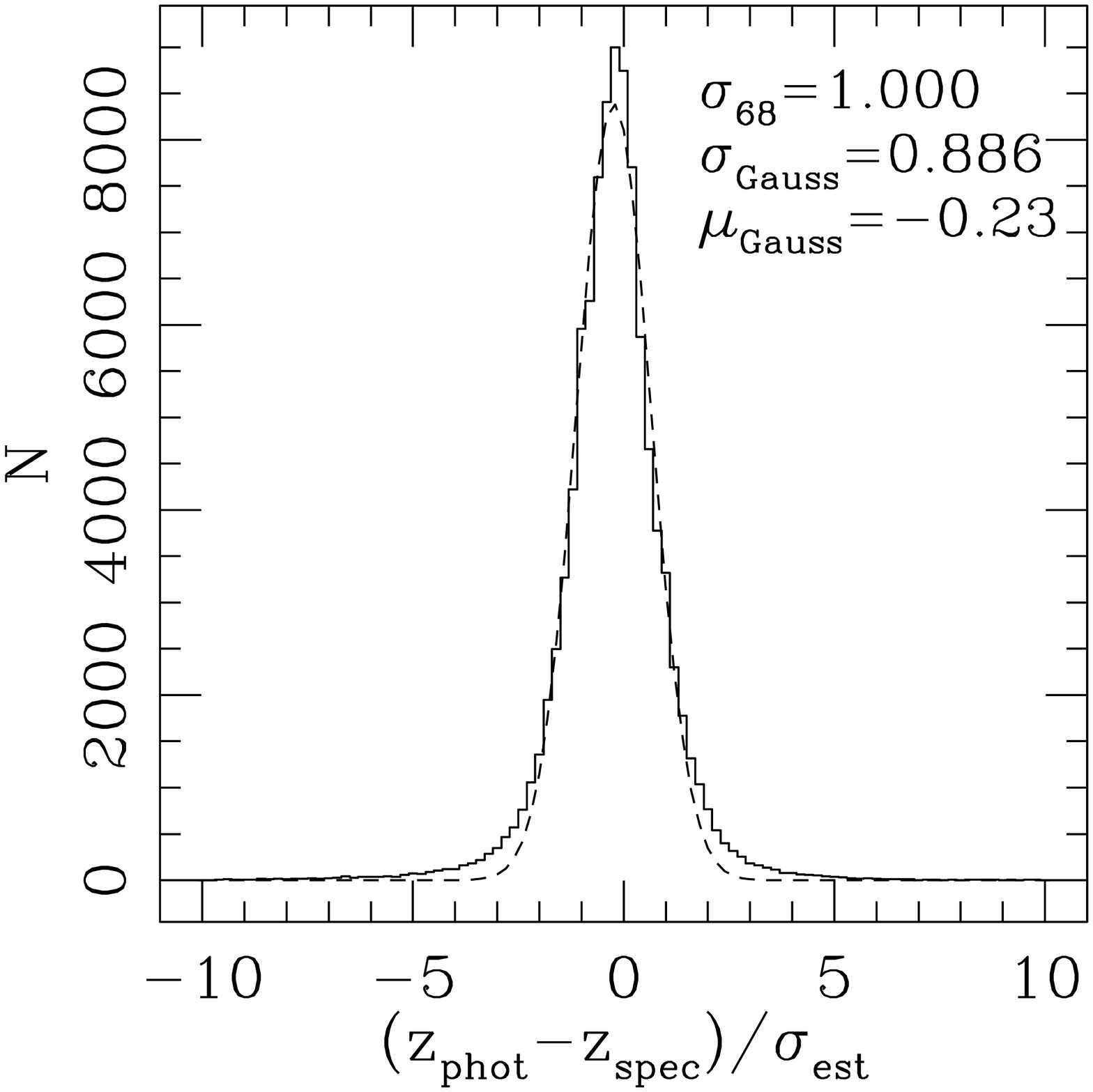}}
    \end{tabular}
  \end{center}
  \caption{{\it Left:} Estimated Error vs.\ empirical error 
for NNE applied to the DES catalog with Hyperz photo-z's.
{\it Right:} Error distribution for the same data.\label{NNEHYPE}}
\end{figure}
 
As noted above, 
the NNE and the Kd-tree error methods can be used in conjunction with any 
photo-z estimator, either training-set or template-based, provided 
there exists a subset of the photometric sample with spectroscopic redshifts. 
As an illustration, we use the Hyperz template fitting method 
to calculate photometric redshifts for the full DES mock catalog (shown in the 
lower panel of Fig. \ref{plot:zpzsDES}).
We then use 50,000 objects from the DES catalog as a training set for NNE 
and calculate photo-z errors for the remaining photometric objects.
Figure \ref{NNEHYPE} shows the estimated vs. empirical error (left panel) and the error distribution 
(right panel) for this example. 
The NNE error estimate works well, though as before it results in an underestimate 
when the errors are very large ($\Delta z>0.25$).
The error distribution is not as well fit by a Gaussian in this case; this is not 
surprising, since the photo-z estimate in this case has a net bias of 
$\sim 23\%$.
However, the error estimator is able to account for the bias and is still 
able to predict the error to within 12\% in $\sigma_{\rm Gauss}$.
This ability to include the bias in the error estimates makes the training 
set error estimate approach particularly powerful compared to methods based
on magnitude error propagation (see \S \ref{md}). 

In our implementation of the NNE, computation of the NNE is expensive 
compared to the Kd-tree method.
In the naive implementation, computation time to find the nearest 
objects scales as $N_{\rm T} N_{\rm P}$, 
where $N_{\rm T}$ and $N_{\rm P}$ are the number of objects in the training set and 
the photometric set, respectively \citep[see, e.g.,][]{pre92}.
In contrast, the Kd-tree method scales as $N_{\rm P}\log N_{\rm T}$.
For most training-set photo-z methods, including the Neural Network, the 
computation time scales as $N_{\rm P}$.
Therefore, for a sizeable training set ($N_{\rm T} \sim 10,000$ objects), 
the NNE computation dominates the time involved in estimating
the photo-z's and their errors.
Fortunately, the method is trivially parallelizable, because the NNE 
calculation of one object in the photometric set is independent of all the 
other objects in the same set.
Taking advantage of this parallelization, the NNE estimator has been 
successfully applied to a data set as 
large as the SDSS DR6 \citep{ade07} containing more than 78 million
galaxies with Neural Net photo-z's.\citep{oya08}.
In addition, tree-structured nearest neighbor search methods, such as
 the Cover-Tree \citep{bey06}, can be used to improve the computation 
time to $O(N_{\rm P}\log N_{\rm T})$, essentially eliminating the difference
between the Kd-tree and NNE methods.

\subsection{Non-representative training set} \label{extrap}
 
\begin{figure}
  \begin{center}
    \begin{tabular}{cccc}
      \resizebox{40mm}{!}{\includegraphics{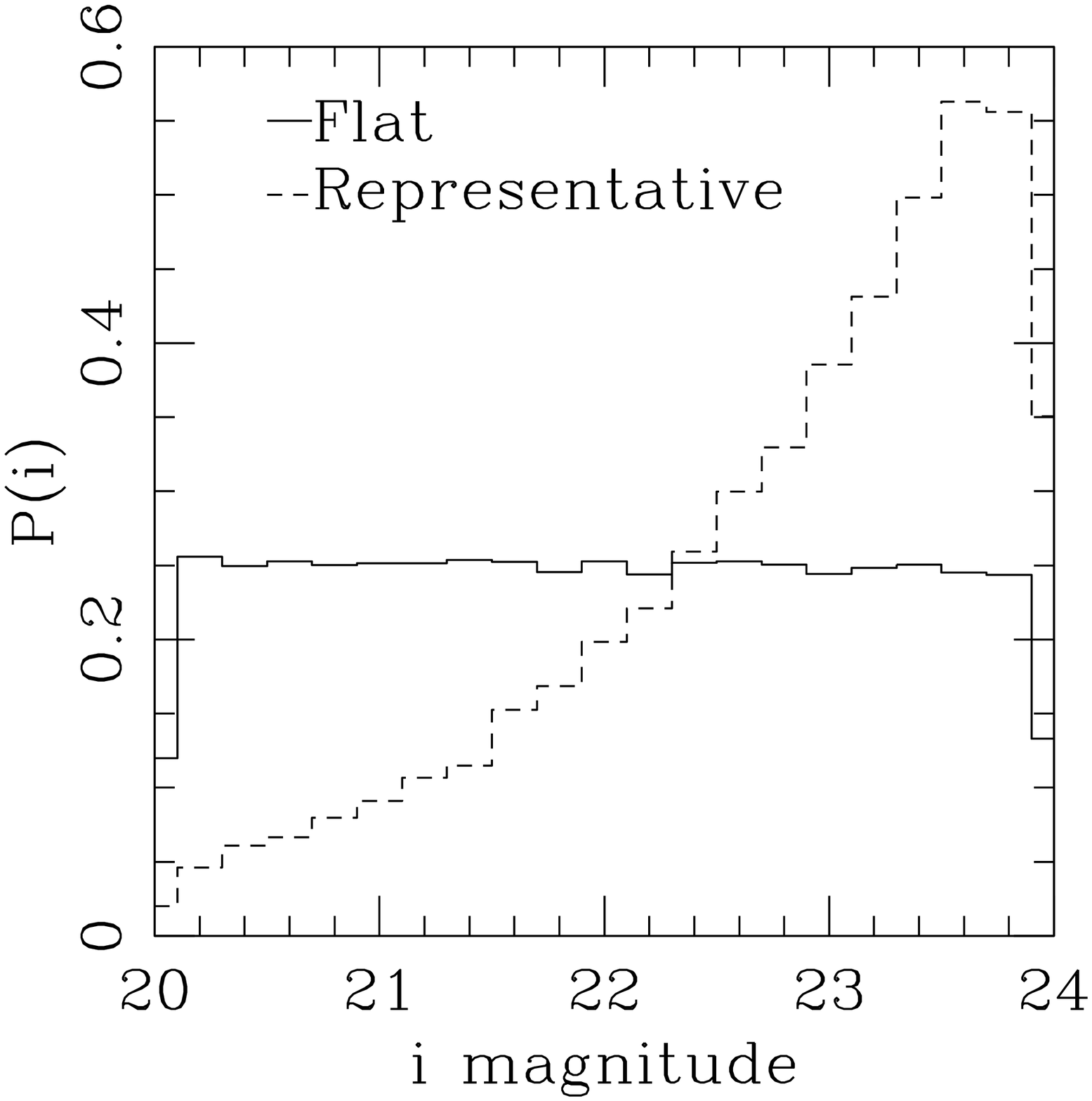}} &
      \resizebox{40mm}{!}{\includegraphics{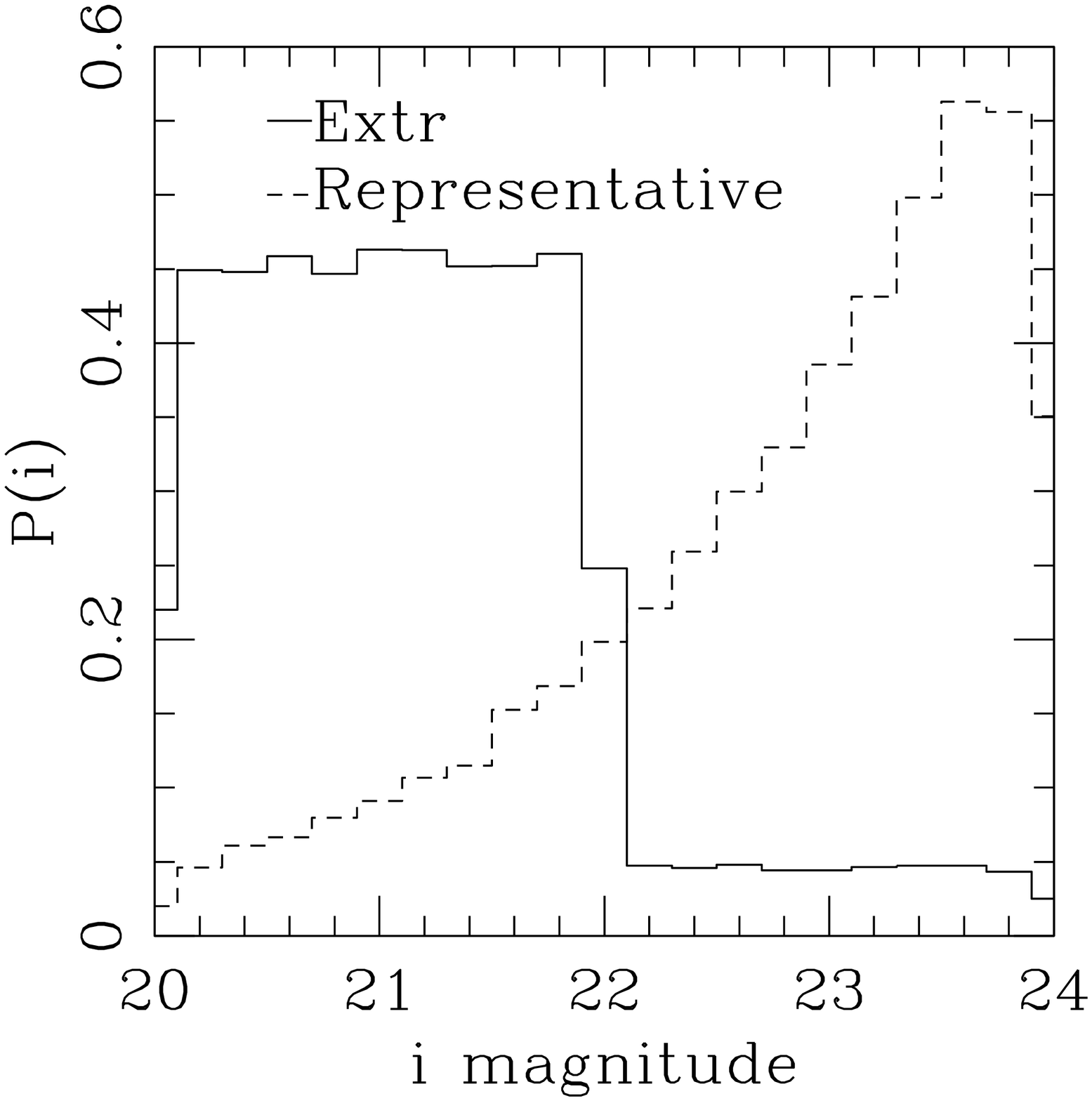}} \\
      \resizebox{40mm}{!}{\includegraphics{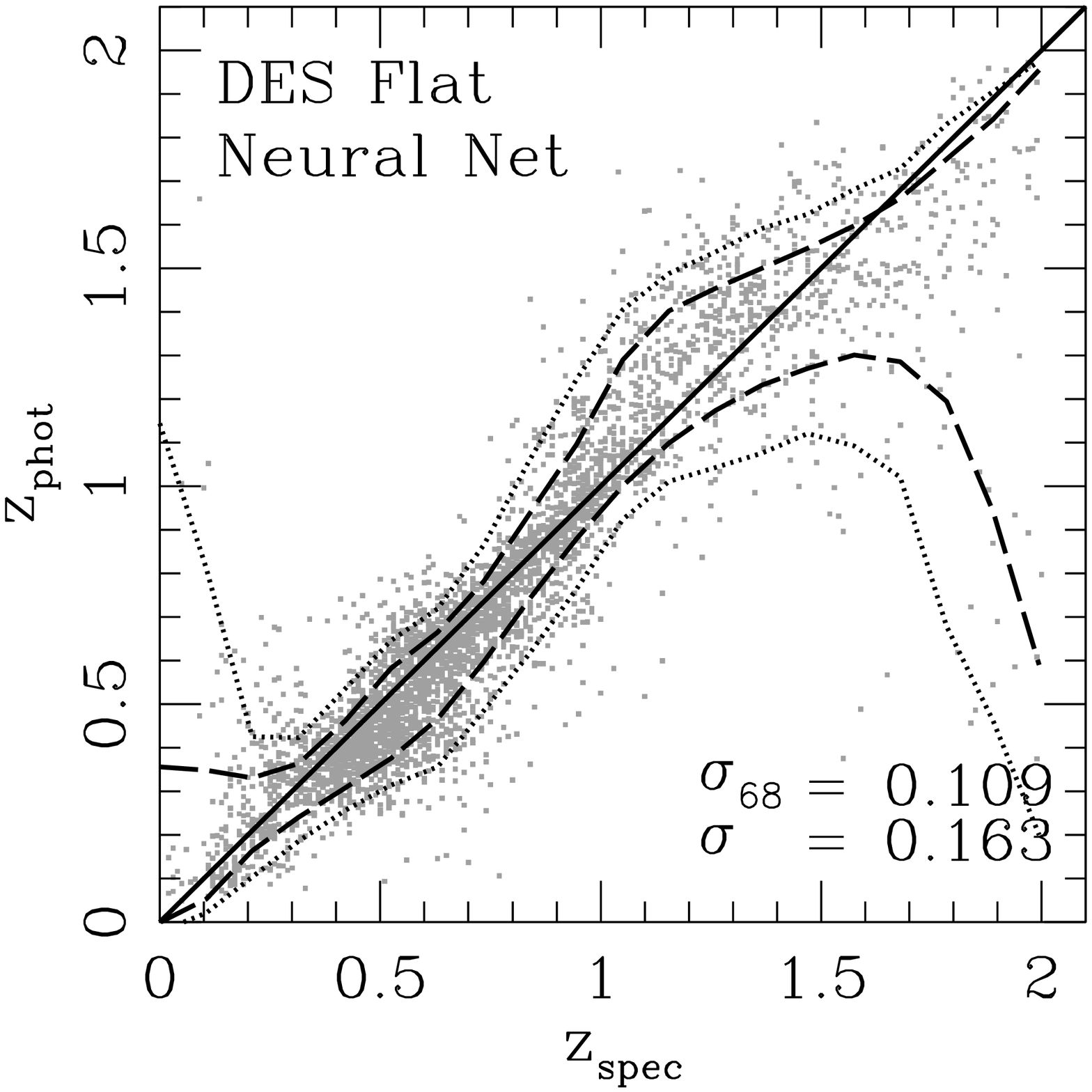}} &
      \resizebox{40mm}{!}{\includegraphics{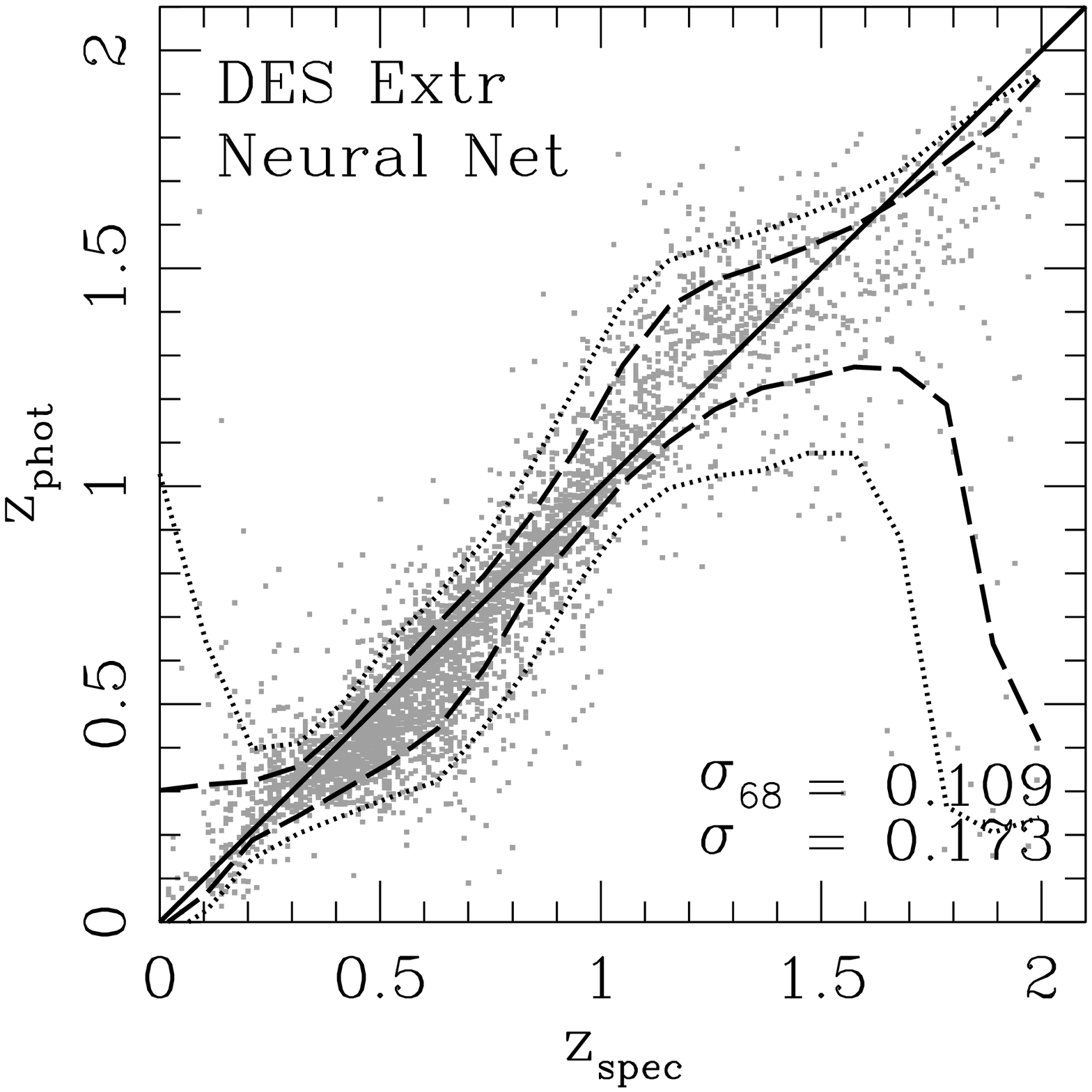}} \\
  \end{tabular}
    \caption{
      {\it Top row:} $i$-magnitude distribution for the Flat and Extr 
      non-representative training sets.  The representative $i$-magnitude
      distribution is plotted in dashed lines for comparison.
      {\it Bottom row:} The Neural Network $z_{\rm phot}$ vs $z_{\rm spec}$ of 
      the DES mock 
      photometric set calculated using DES Flat training set ({\it left panel}) 
      and DES Extr training set ({\it right panel}).
    }
    \label{plot:nonrep}
  \end{center}
\end{figure} 

\begin{figure}
  \begin{center}
    \begin{tabular}{cccc}
      \resizebox{40mm}{!}{\includegraphics{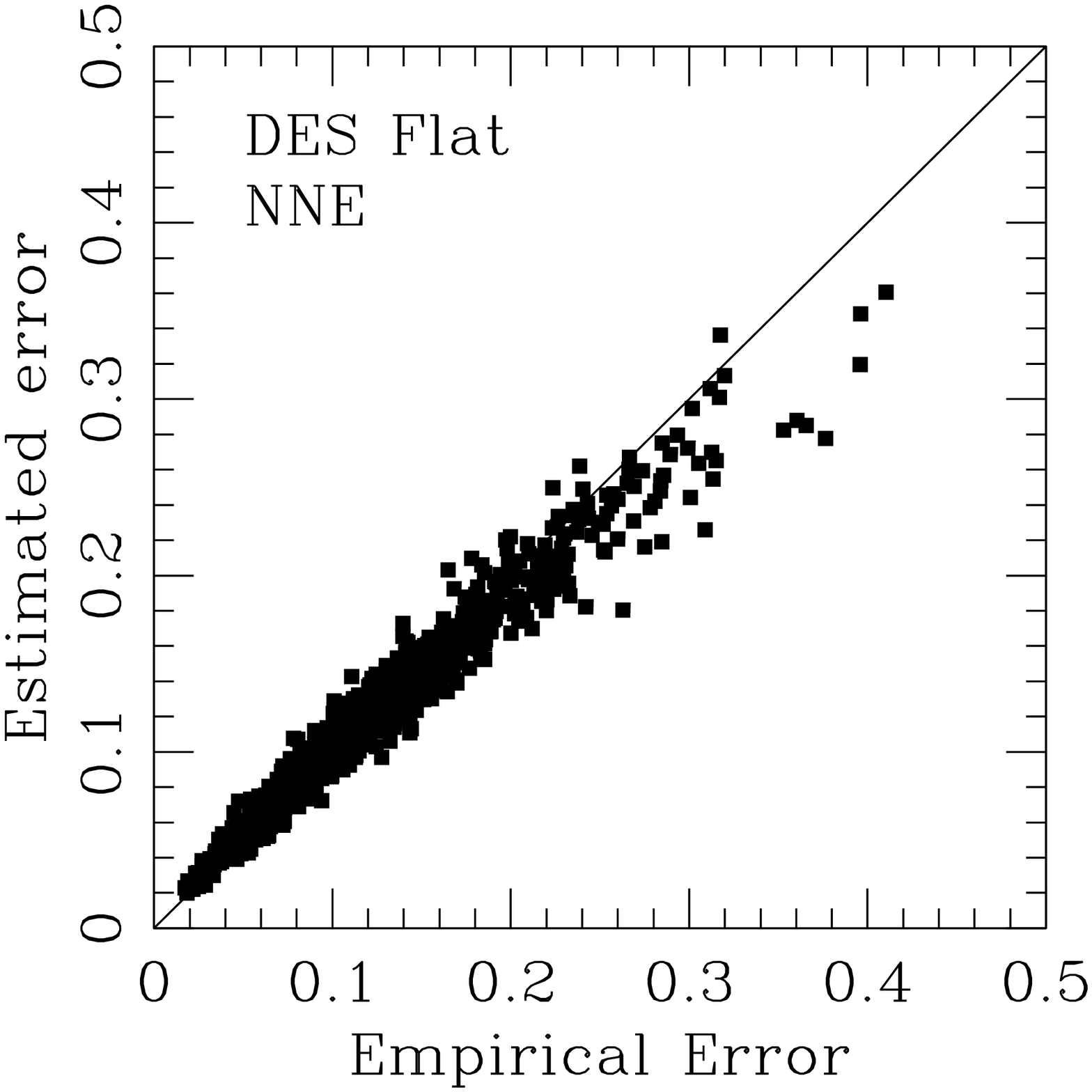}} &
      \resizebox{40mm}{!}{\includegraphics{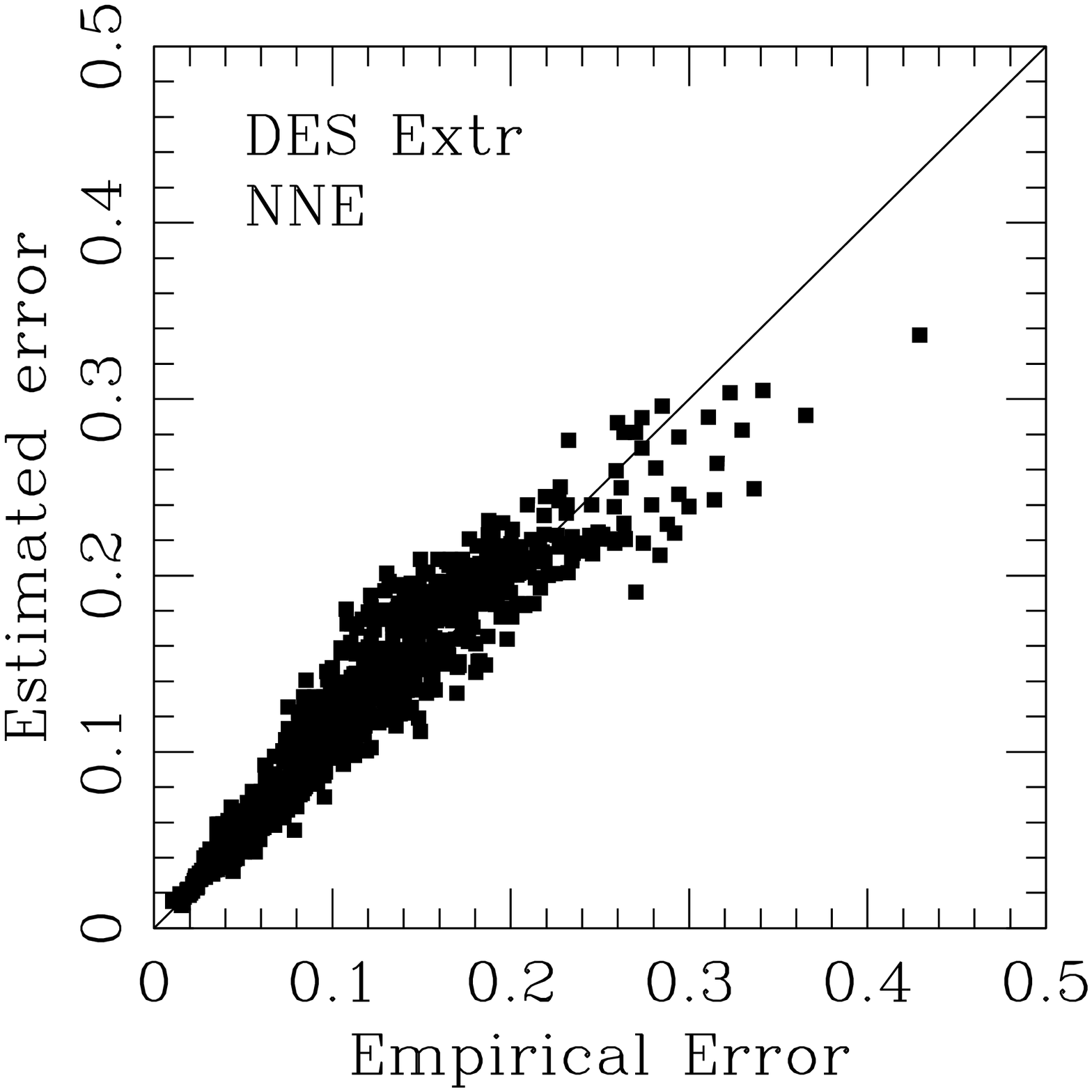}} \\
      \resizebox{40mm}{!}{\includegraphics{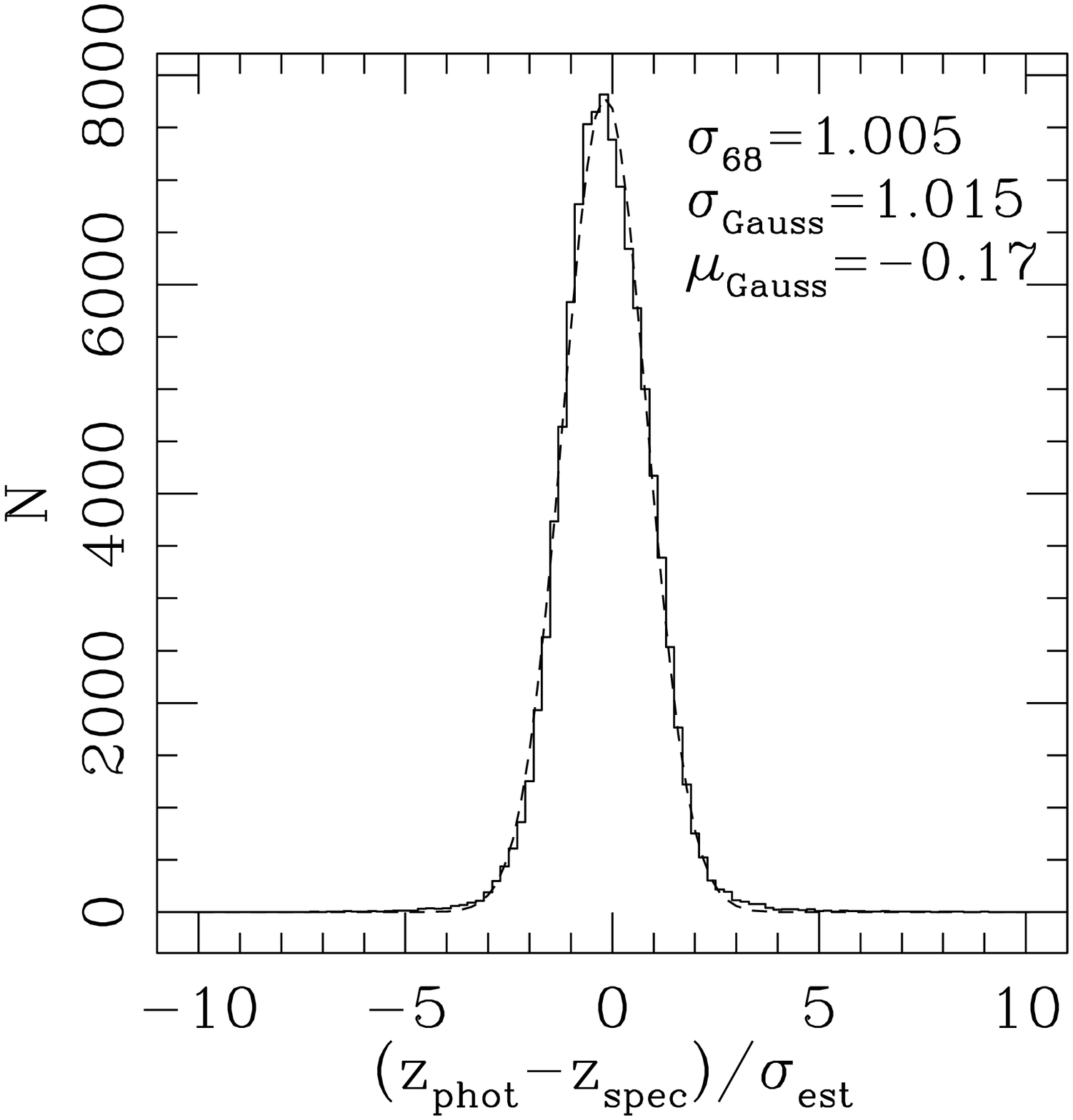}} &
      \resizebox{40mm}{!}{\includegraphics{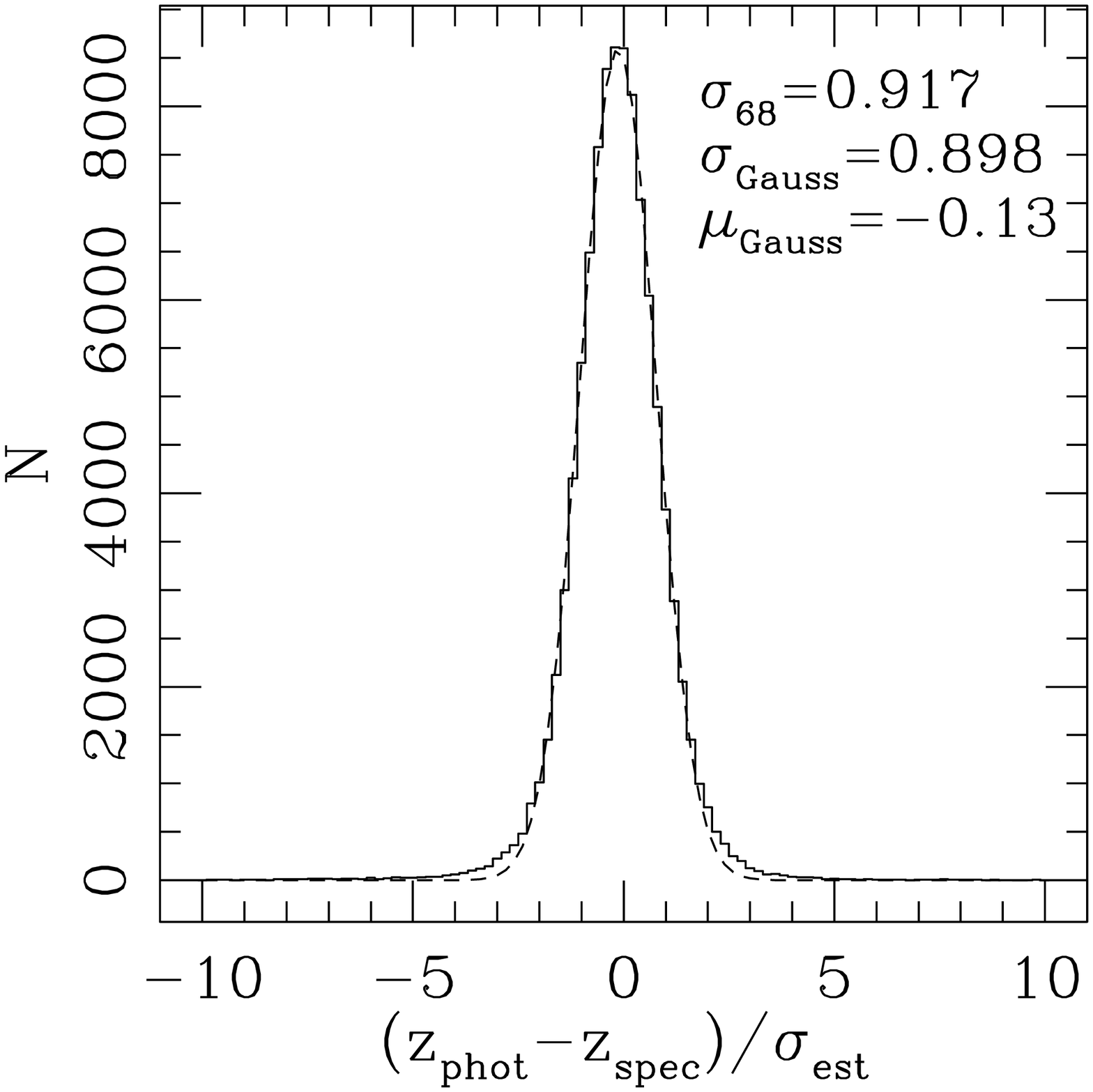}} \\
  \end{tabular}
    \caption{
      {\it Top row:} NNE error vs empirical error calculated using
      two non-representative training sets.
      {\it Bottom row:} Error residual distributions.
    }
    \label{extrapplot}
  \end{center}
\end{figure} 

The training-set based error estimators we have introduced 
rely on the spectroscopic training set to 
characterize the errors of the photometric set. 
Hence, the quality of the error estimate depends in principle 
on the degree to which the training set 
is a representative subsample of the photometric set. Since spectroscopic 
samples often are {\it not} simply random subsets of the parent photometric 
samples from which they are drawn, one might have concerns about the 
robustness of these error estimates. Here, we consider cases of 
non-representative training sets and 
show that the training-set error estimators perform satisfactorily provided 
the training set covers the full magnitude range of the photometric sample.

In order to illustrate the issue, we have constructed two non-representative
training sets using the DES catalog generator.
One training set (labelled Flat) has a flat $i$-magnitude distribution at $i<24$ instead of the 
increasing distribution characteristic of a flux-limited sample; bright (faint) 
objects are over- (under-)represented compared to the photometric sample. 
The second training set (labelled Extr) has an  
$i$-magnitude distribution highly skewed to bright magnitudes, 
$i < 22$, since a spectroscopic set typically does not go as faint as 
the corresponding photometric sample. 
Both training sets have flat redshift and SED type distributions, differing
from those of the fiducial DES mock catalog.
The $i$-magnitude distributions, as well as the $z_{\rm phot}$ vs $z_{\rm spec}$
plots, are shown in Figure \ref{plot:nonrep}.
Each training set contains 50,000 galaxies.
We used the training sets to derive Neural Network photo-z solutions, 
which were then used to estimate photo-z's for the DES 
mock photometric catalog. 
Photo-z errors were estimated using the NNE method, again using the same 
non-representative training sets in each case.
In Fig. \ref{extrapplot}, 
we show the estimated vs. empirical error (top panels) and the 
error distributions (bottom panels) for the 
two cases.
We see that the NNE error method estimates the errors correctly at the 
$\sim 10\%$ level while maintaining Gaussianity in both cases.
In the case of the Flat training set, the error accuracy degradation is 
less than 1\% compared to the representative training case. 
Given the fact that the Neural Network photo-z quality is itself 
degraded by $\sim10\%$ compared to the representative case in scatter,
these results show that the NNE error estimator is robust against 
differing distributions of the training and photometric sets.

A possible approach to the issue of non-representative training sets 
would be to resample
or weight the training-set objects to obtain a distribution 
that matches the distribution of 
photometric observables (magnitudes, colors, etc.) of the photometric sample. 
In the case of the DES catalog and the two non-representative training
sets used above, this resampling results in a marginal 
improvement in the error estimate at the $\sim2\%$ level in both
$\sigma_{68}$ and $\sigma_{\rm Gauss}$. 
We plan to offer further discussions and test results in subsequent 
articles, currently in preparation \citep{lim08,cun08}.

\section{Comparison with other error estimators} \label{section:comparison}
Other photo-z error estimators have been proposed in the literature.
Two commonly used estimators are the $\chi^{2}$ error in
template fitting methods, such as Hyperz \citep{bol00}, and the 
propagation of magnitude errors that is found in, for example, 
ANNz \citep{col04}.
In this section, we discuss the performance of these error estimators
and consider the advantages and disadvantages of our training-set based
error estimators compared to these methods.

\subsection{$\chi^{2}$ error estimate}
\begin{figure}
  \begin{center}
    \begin{tabular}{cc}
      \resizebox{40mm}{!}{\includegraphics[angle=0]{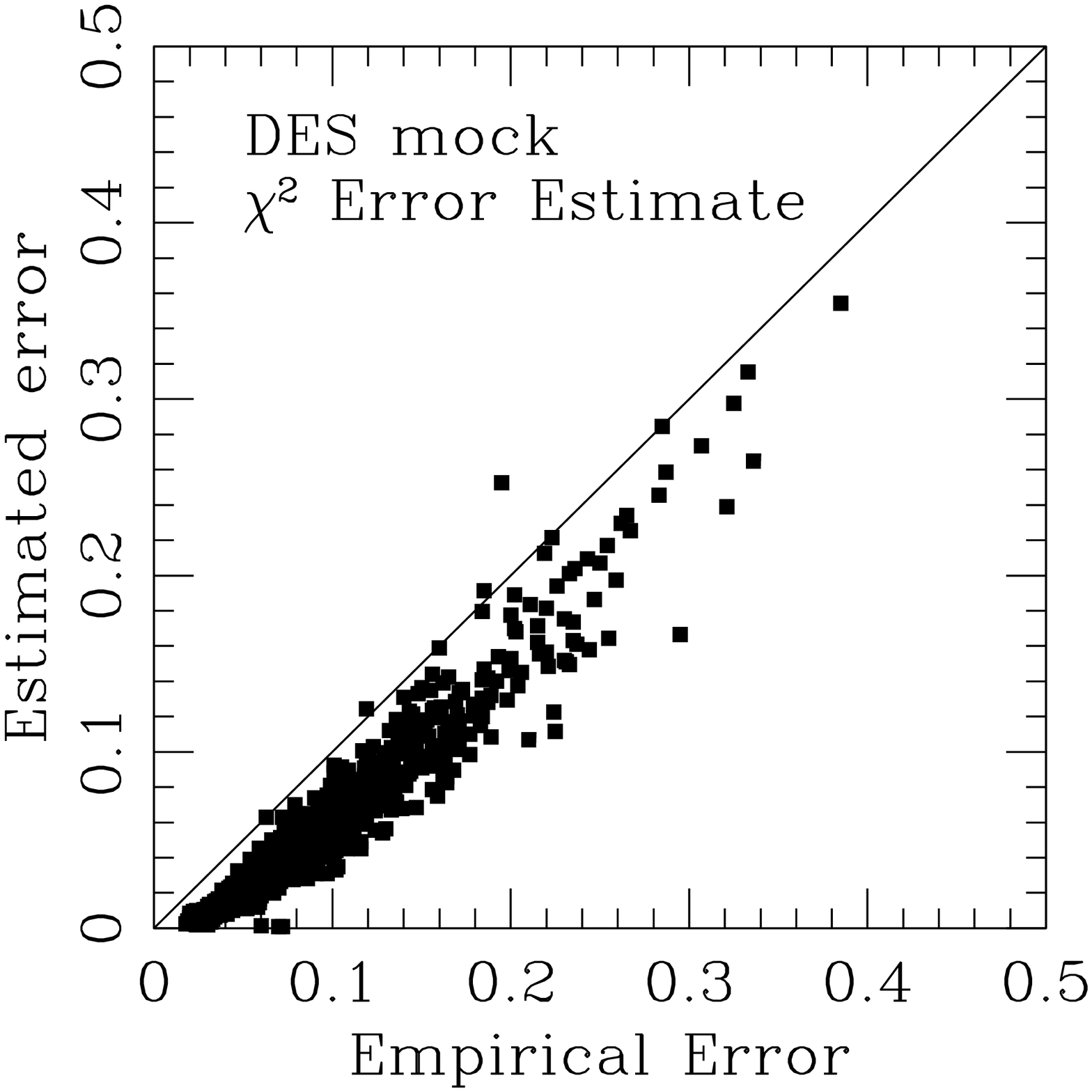}} &
      \resizebox{40mm}{!}{\includegraphics[angle=0]{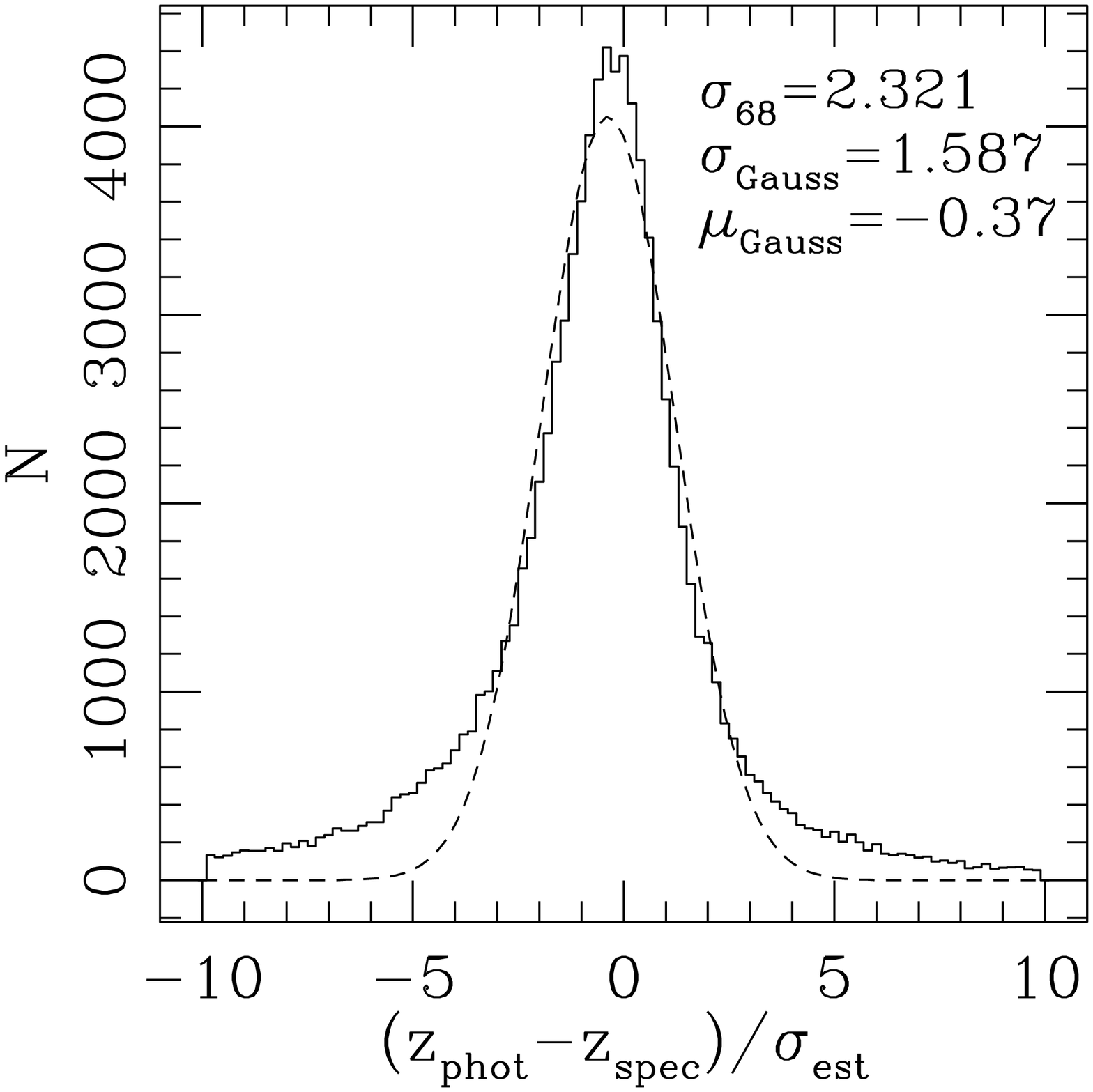}} \\
    \end{tabular}
    \caption{{\it Left}: $\chi^{2}$ estimated error vs empirical error 
      for the DES mock catalog, using the Hyperz photo-z estimator.
      {\it Right:}  $\chi^2$ error residual distribution, along with
      Gaussian fit. For the comparable training-set 
      error, see Fig. \ref{NNEHYPE}.
    }
    \label{plot:chisq}
  \end{center} 
\end{figure}

Template-fitting photo-z methods often use $\chi^2$ minimization to determine 
the best-fit {\zphot} and spectral type. The quantity to be minimized is 
\begin{equation}\label{chieq}
\chi^2=\sum_{k=1}^{N_{\rm m}}\left[\frac{F_{\rm obs}^{k}-a\cdot F_{\rm temp}^{k}(z)}{\sigma_{F}^{k}}\right]^2, \label{chi1}
\end{equation}
\noindent where $F_{\rm obs}^{k}$ is the observed flux in passband $k$,
 $\sigma_{\rm F}^{k}$ is the 
corresponding uncertainty in the flux, $F_{\rm temp}^{k}(z)$ is the 
flux of a template SED redshifted to a 
given $z$, $a$ is a normalization factor, and $N_{\rm m}$ is the number
 of passbands in which measurements are available. This statistic is minimized over 
redshift and over the set of template SEDs.

When a model being fit to data is linear in the fit parameters, 
the probability distribution for the $\chi^2$ statistic is 
the chi-square probability distribution 
for $\nu$ degrees 
of freedom, $P(\chi^2|\nu)$ \citep{pre92}.
Given the value of $\chi^2=\chi^2_{\rm min}$ that minimizes Eq. 
\ref{chieq}, the corresponding $P(\chi_{\rm min}^2|\nu)$
gives the probability that the observed $\chi^2$ for a correct 
model should be less than $\chi^2_{\rm min}$. This probability 
can be used to calculate redshift 
confidence intervals. 
Given a confidence level $\alpha$ ($0<\alpha<1$), define the quantity 
$\Delta_{\chi^2}$
such that \citep{avn76} 
\begin{equation}\label{deltchieq}
P(\chi^2  \leq \Delta_{\chi^2}|\nu)=\alpha.
\end{equation}
\noindent The level-$\alpha$ $z_{\rm phot}$ confidence interval is given by the set of 
all redshifts for which
\begin{equation}\label{confchieq}
\chi^2(z) - \chi^2_{\rm min} \leq \Delta_{\chi^2}~,
\end{equation}
\noindent where $\chi^2(z)$ is minimized over spectral type and the coefficient $a$. 
That is, $\Delta_{\chi^2}$ is simply the increment in $\chi^2$ required to
cover the region of parameter space with redshift confidence $\alpha$.
Here, we are interested in comparing the 68\% confidence interval of 
the photometric redshift, so we set the parameter $\alpha = 0.68$.

In Figure \ref{plot:chisq}, we show the $\chi^2$-estimated error vs. empirical 
error and the residual error distribution for the Hyperz photo-z estimator 
applied to the DES mock catalog.
The $\chi^{2}$ error underestimates the true error by about
a factor of two.
Furthermore, the distribution of the error residual divided by the 
estimated error is decidedly non-Gaussian, exhibiting strong tails.
We attribute the underestimate to the fact that the chi-square 
distribution is not a realistic description of the true photo-z error 
distribution,
given the relatively strong degeneracies present in the catalog.
In a test using an artificial mock catalog containing only early-type galaxies, 
in which the degeneracy between redshift and galaxy SED type is removed, 
we found that the $\chi^2$ estimator was accurate at the $\sim 30$\% level.
In addition, the model used in the $\chi^{2}$ error estimator assumes 
that the fitting function, $F_{\rm temp}^{k}(z)$, is linear in the 
fitting parameters, namely the redshift.
In reality, the template fitting functions are highly non-linear, and
therefore it is not surprising that the $\chi^{2}$ error estimator does
not robustly predict the correct errors. 

We also tried to compute $\chi^{2}$ errors for Hyperz applied to the SDSS 
catalog, 
but we were not able to obtain sensible estimates.  
We found no discernible correlation between the $\chi^{2}$ errors and 
the true errors of the photo-z estimate.
We discuss this issue further at the end of section \ref{md}.

\subsection{Error estimate from Magnitude Derivative (MDE)}\label{md}
\begin{figure}
  \begin{center}
    \begin{tabular}{cc}
      \resizebox{40mm}{!}{\includegraphics[angle=0]{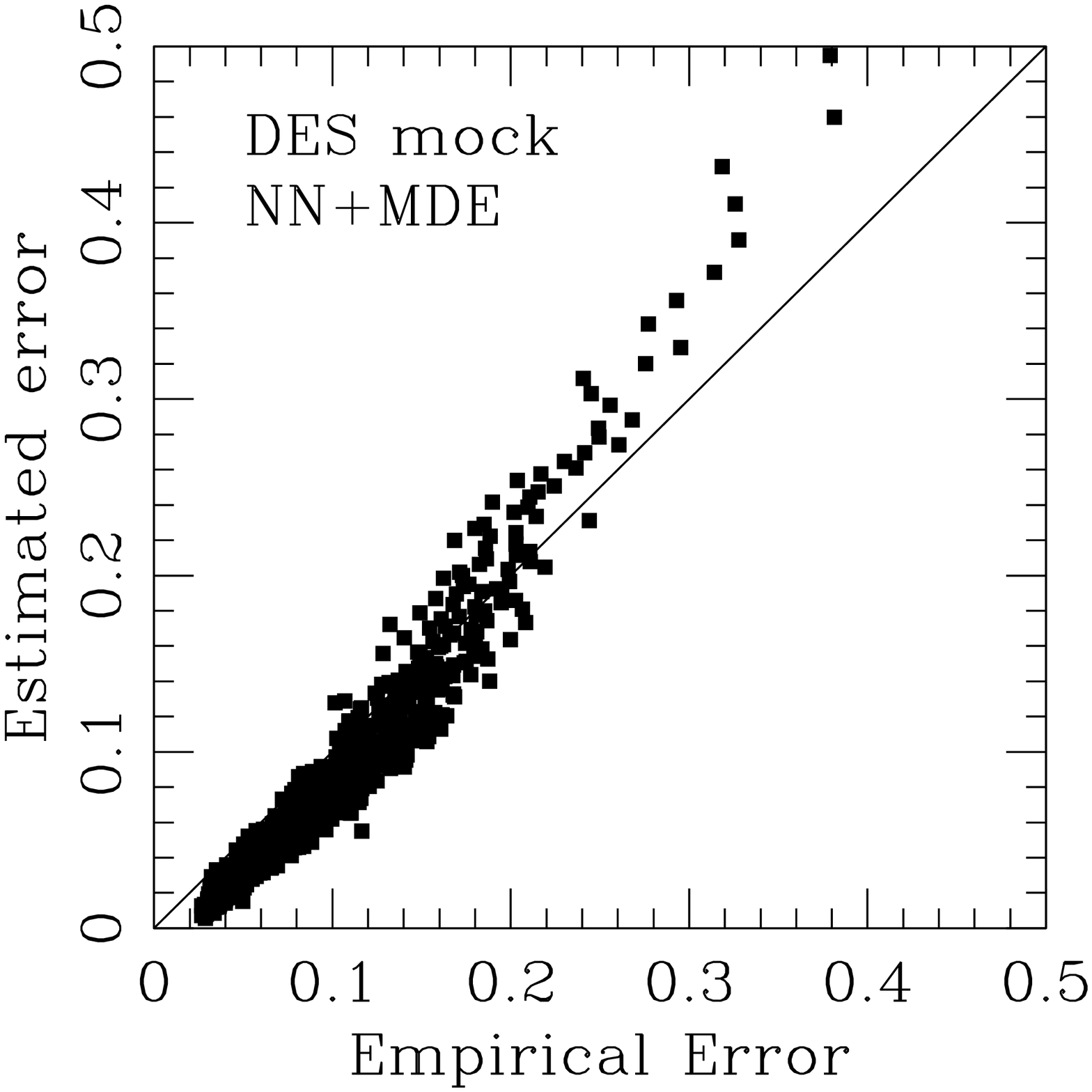}} &
      \resizebox{40mm}{!}{\includegraphics[angle=0]{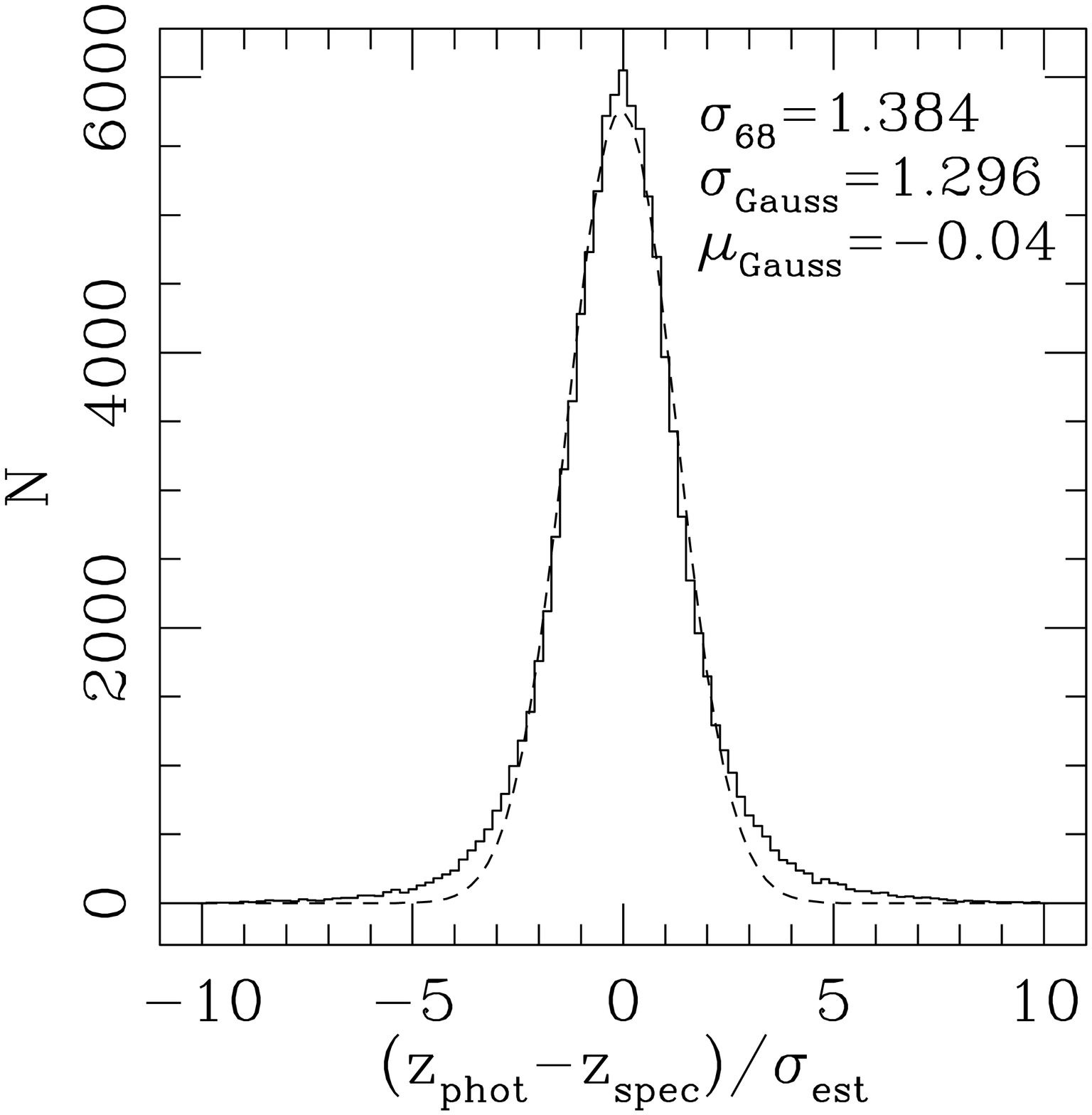}} \\
    \end{tabular}
    \caption{
      {\it Left}: MDE error vs empirical error for the DES mock catalog, using the NN photo-z estimate.
      {\it Right}: Error residual distribution for MDE error for the DES mock catalog. For 
comparison with the training-set error estimators, see Figs. \ref{train1}a,c.
    }
    \label{plot:mde}
  \end{center} 
\end{figure}

The basic assumption underlying photo-z estimates is that there is a one-to-one 
mapping from photometric observables, e.g., magnitudes, to redshift. In training-set 
photo-z methods, this mapping is given by an explicit, usually analytic, function 
of magnitudes $m^{\mu}$ and fit coefficients $c_{\rm k}$,  
\begin{equation}
z_{\rm phot} = z_{\rm phot}(c_{\rm k}, m^{\mu}) ~,
\end{equation}
where the $c_{\rm k}$ are determined from the spectroscopic training set by minimizing 
a {\it score function}, a measure of the error residuals of the photo-z estimates. 
To first order, we can propagate the 
coefficient errors $\sigma_{c_{k}}$ and the magnitude errors $\sigma_{m}$
to the photo-z errors $\sigma_{z}$ as
\begin{equation}
\sigma_{z}^2 = \sum_{k=1}^{N_c} \left(\frac{\partial z_{\rm phot}}{\partial c_{k}}\right)^2 \sigma_{c_{k}}^2
+\sum_{\mu=1}^{N_m} \left(\frac{\partial z_{\rm phot}}{\partial m^{\mu}}\right)^2 \sigma_{m^{\mu}}^2 ~.
\label{coefmagdef}
\end{equation} 
\noindent If the training set is sufficiently large
(say, $\sim 10,000$ objects), the photo-z errors due to errors in the 
model fit coefficients are typically negligible compared to 
those arising from magnitude error propagation. 
Therefore we will concentrate on the latter  
and define the Magnitude Derivative 
error (MDE) as the second term in Eqn. \ref{coefmagdef} \citep{col04}.
For polynomial fitting and NN photo-z methods, analytic 
expressions for the derivatives (see, e.g., \citet{bis95} for the 
case of NN) can be used. However, 
we may also calculate these derivatives by finite difference, in which 
case MDE can be applied to any photo-z estimation method, including template fits.

Figure \ref{plot:mde} shows the performance of the MDE error calculation for the
DES mock catalog using neural network photo-z's. 
MDE errors underestimate the true error by approximately
40\% for this case. Although the 
error residuals are nearly Gaussian, 
the tails of the error distribution are more pronounced than the
tails for the NNE error, signaling the failure of MDE to correctly identify
catastrophic photo-z errors.

\citet{col04} identify a second source of error in neural network photo-z's: in the 
training process, the score function 
typically has many local minima with similar values. As a result, 
networks that start the minimization process at different initial values 
for the fit coefficients can 
end up in different local minima, resulting in slightly different photo-z
 estimates for the same input magnitudes. 
The variance in photo-z 
estimates due to this effect is an additional contribution to the photo-z
 error.
By retraining our networks with different initial conditions, 
we find that the contribution of such an effect to the photo-z error is 
small ($<1\%$ of MDE) for our
two catalogs, not enough to account for the underestimate of the 
MDE errors when applied to neural network photo-z estimates.

The $\chi^2$ and MDE error estimators are both predicated on the accuracy of 
the quoted magnitude errors. However, photometric errors are often 
difficult to estimate accurately \citep[e.g.,][]{scr05}.  
The problem is further exacerbated if the magnitude errors in different passbands are 
correlated with each other, thereby violating the assumptions made
in the  $\chi^{2}$ fit and in magnitude error propagation. 
Because of these difficulties, the MDE errors applied to the NN photo-z 
estimates for the SDSS catalog are only 
weakly correlated with the true errors, 
similar to the case of $\chi^{2}$ error applied to the SDSS.
A key advantage of the training-set based error estimators is 
that they do not depend on the measured magnitude errors.

\section{Reducing Catastrophic Outliers: Culling objects by estimated error}\label{section:culling}

\begin{figure}
  \begin{center}
    \begin{minipage}[t]{85mm}
      \resizebox{85mm}{!}{\includegraphics[angle=0]{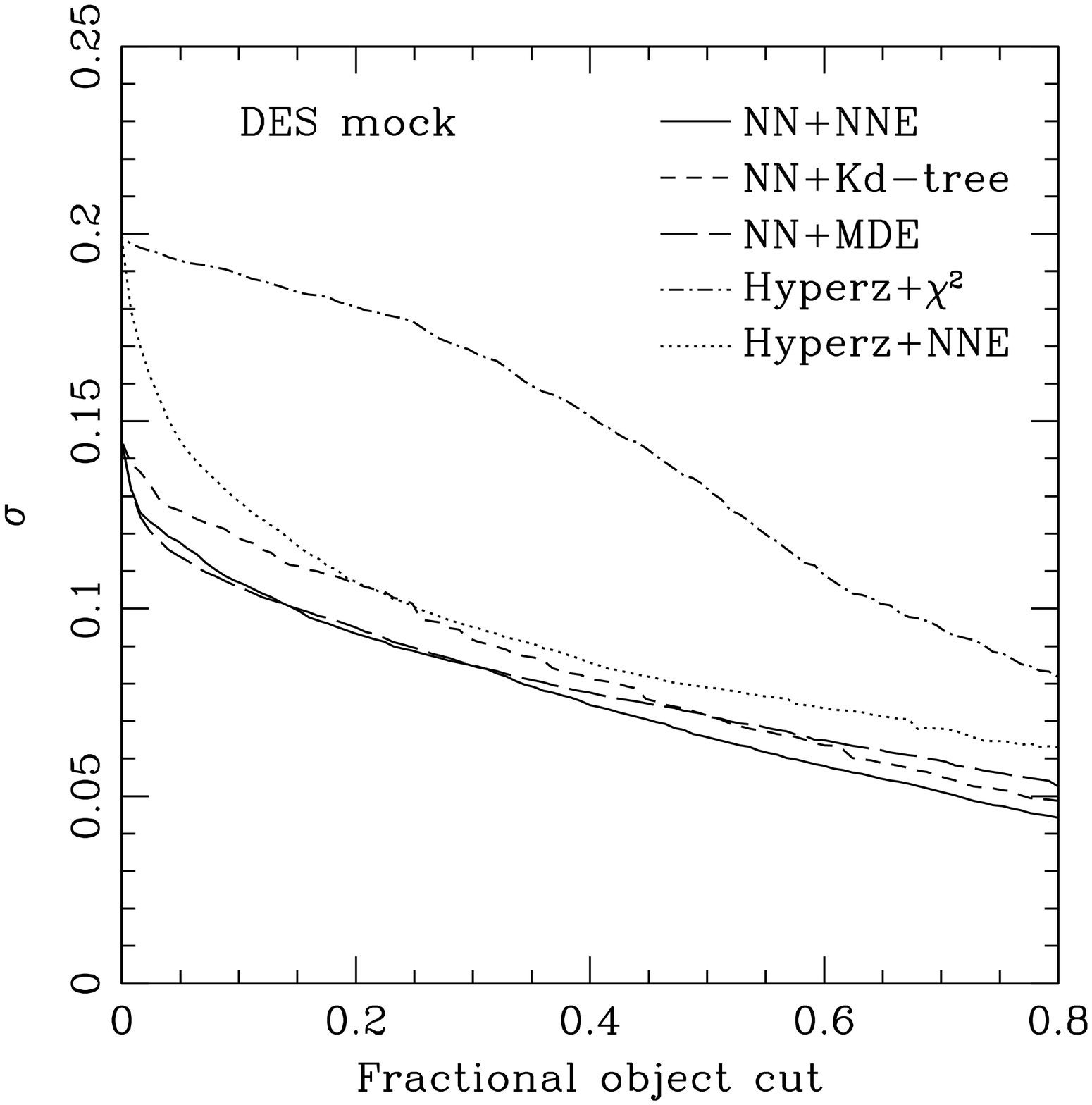}}
    \end{minipage}
    \begin{minipage}[t]{85mm}
      \resizebox{85mm}{!}{\includegraphics[angle=0]{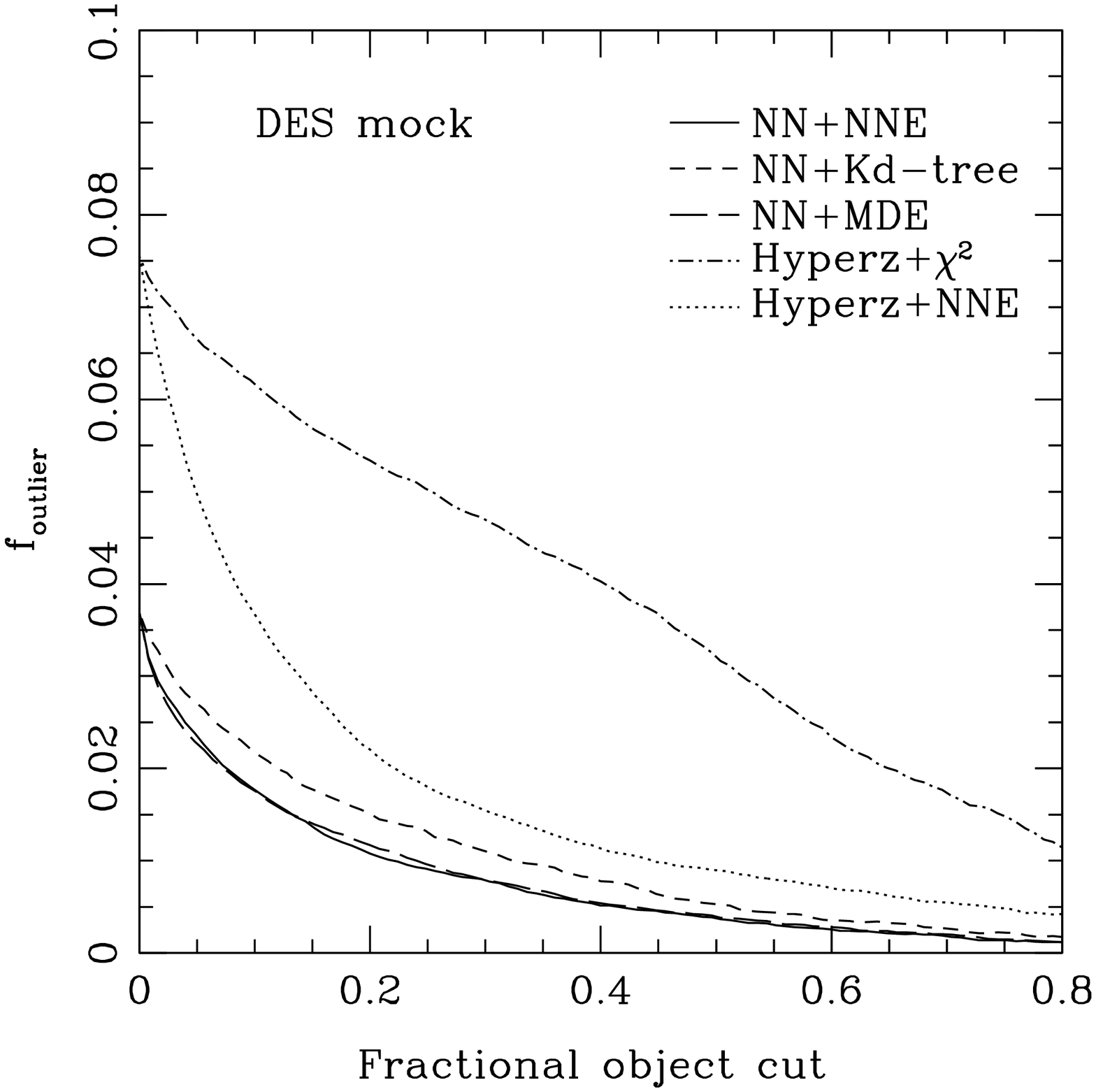}}
    \end{minipage}
    \caption{{\it Top}: Reduction in photo-z scatter $\sigma$ when objects with 
large estimated photo-z errors are culled from the sample, using two photo-z 
estimators, NN and Hyperz, and four error estimators, NNE, Kd-tree, MDE, and $\chi^2$. 
Horizontal axis is the the fraction of objects 
      culled from the DES catalog. 
      {\it Bottom}: Reduction in outlier fraction when objects are culled by 
      estimated photo-z error. For the DES catalog, 
      the outlier fraction is defined as the fraction of objects with
      $|z_{\rm phot} - z_{\rm spec}| > 0.3$.
    }
    \label{plot:cutDES}
  \end{center}
\end{figure} 

\begin{figure}
  \begin{center}
    \begin{minipage}[t]{85mm}
      \resizebox{85mm}{!}{\includegraphics[angle=0]{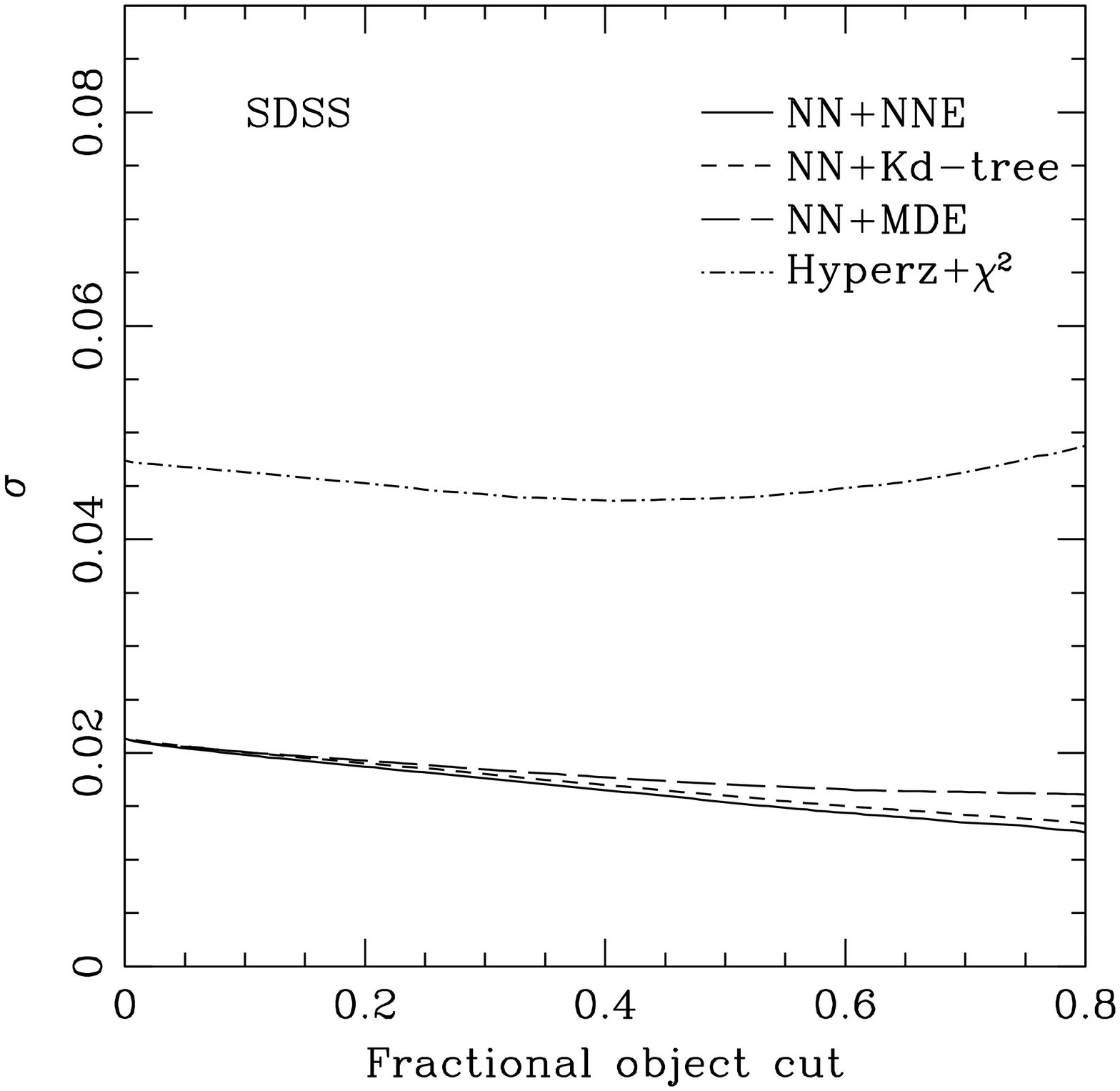}}
    \end{minipage}
    \begin{minipage}[t]{85mm}
      \resizebox{85mm}{!}{\includegraphics[angle=0]{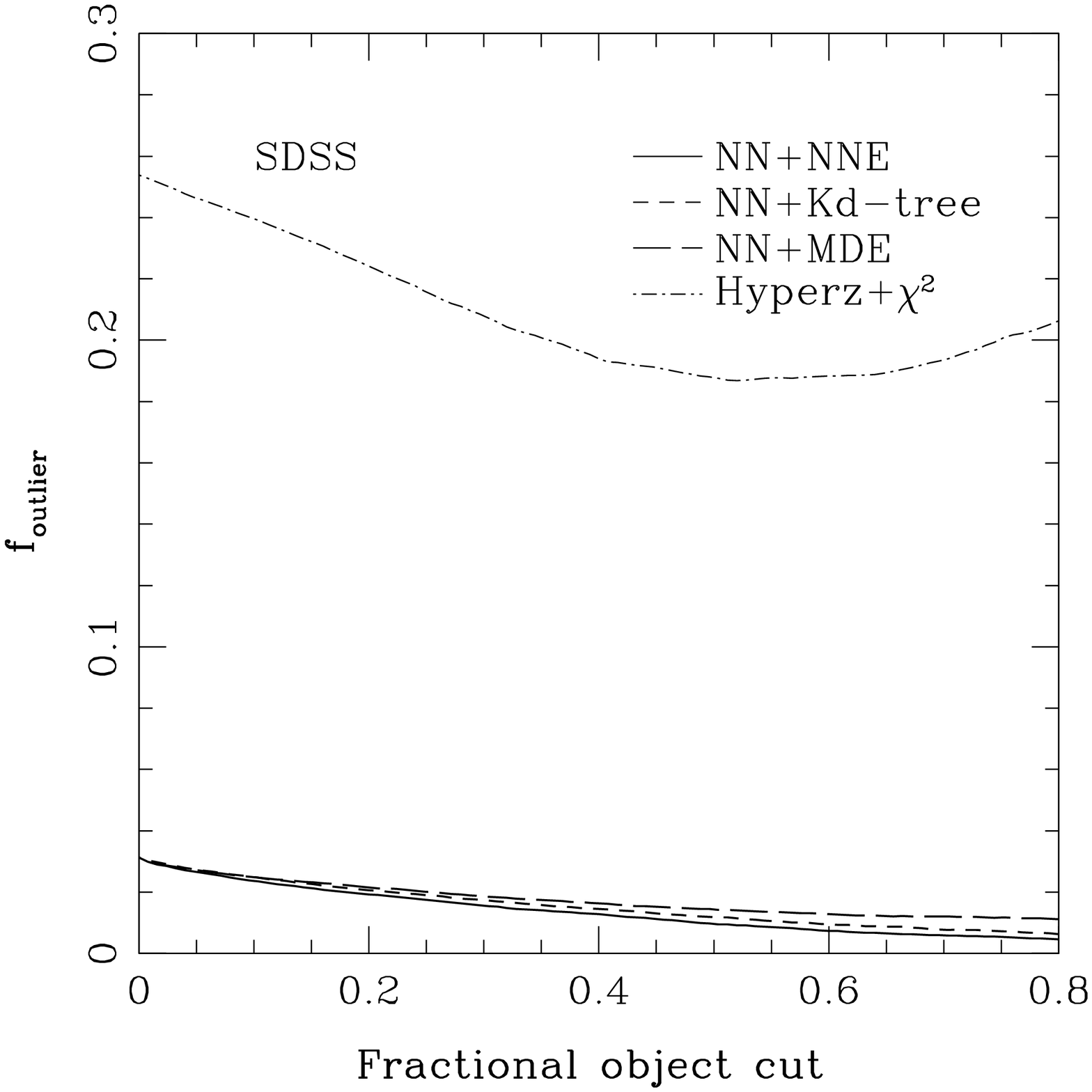}}
    \end{minipage}
    \caption{Same as Fig. \ref{plot:cutDES}, but for the SDSS catalog. Here the 
      outlier fraction is the fraction of objects with
      $|z_{\rm phot} - z_{\rm spec}| > 0.05$.
    }
    \label{plot:cutSDSS}
  \end{center}
\end{figure} 

In certain analyses, one would like to remove objects with very erroneous, 
so called catastrophic, photo-z estimates from a sample. If the estimated 
photo-z errors are 
reliable, then objects with large estimated errors can be used to 
identify catastrophic photo-z failures. Removing such objects from a sample 
can reduce the scatter and bias in photo-z estimates.

In this study, we define objects with catastrophic errors as those for which 
$|z_{\rm phot} - z_{\rm spec}|$ is large compared to the photo-z scatter, $\sigma$.
Specifically, we define catastrophic errors to be $|z_{\rm phot} - z_{\rm spec}| > 0.3$ 
for the DES catalog 
and $|z_{\rm phot} - z_{\rm spec}| > 0.05$ for the SDSS, corresponding to 
approximately 2.5 times the scatter  for the NN photo-z estimate for each survey.
We define the outlier fraction to be the fraction of objects in a photometric 
sample with catastrophic errors.
We sort the photometric catalogs by the galaxies' 
estimated photo-z errors and track the changes in $\sigma$ and in the outlier fraction 
as we successively remove objects with smaller and smaller estimated error.

In Figure \ref{plot:cutDES}, we show the dependence of the photo-z scatter, $\sigma$, 
and the outlier fraction on 
the fraction of objects culled from the sample based on the estimated error. 
We show results for the four different error estimators
described above (Kd-tree, NNE, $\chi^2$, and MDE) for the DES mock catalog. 
We clearly see that the NNE and the MDE estimators perform the best in 
reducing scatter and outliers, while the $\chi^{2}$
method fails to adequately separate catastrophic photo-z's from the well
behaved ones.
Note that the relatively poor performance of the $\chi^{2}$ method is not 
due to the fact that the Hyperz photo-z scatter is larger: the 
NNE error estimate with the Hyperz photo-z performs significantly better.

Figure \ref{plot:cutSDSS} shows the photo-z scatter and outlier fraction 
for the SDSS catalog. For this case, 
MDE and $\chi^{2}$ do not perform as well in reducing scatter and outliers.  
These error estimators rely on the reported magnitude errors, and as noted above 
the latter are highly correlated between passbands and are non-Gaussian   
for the SDSS catalog.
In fact, culling objects with high $\chi^{2}$ error results in no 
improvement of the scatter, a reflection of the fact that the $\chi^{2}$ error for
the SDSS catalog is not correlated with the actual error of 
the Hyperz photo-z estimates.

\begin{figure}
  \begin{center}
    \begin{center}
      \resizebox{85mm}{!}{\includegraphics[angle=0]{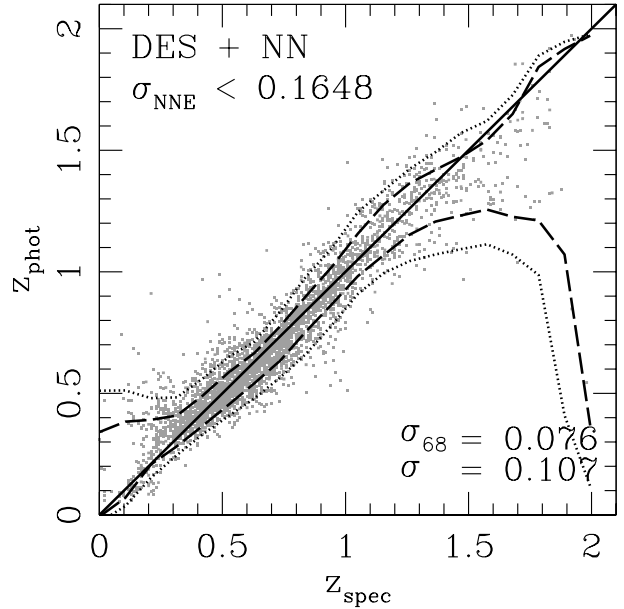}}
    \end{center}
    \caption{{\zphot} vs {\zspec} for the DES catalog when the 10\% of the
      objects with largest estimated NNE error (NNE error larger than 0.1648) have 
      been removed. For comparison, see Figure \ref{plot:zpzsDES}.
      \label{zpzs1}}
  \end{center}
\end{figure}

Figure \ref{zpzs1} shows {\zphot} vs.\ {\zspec} for the DES catalog with NN 
photo-z's when 10\% of the objects, those with the largest estimated NNE errors, 
have been removed. Comparing to the results in the upper panel of Fig. \ref{plot:zpzsDES}, 
this process reduces the photo-z scatter in the remaining objects by $\sim $ 23\%.  
Moreover, most of the catastrophic objects 
 at low redshift are removed, improving the bias and the scatter at 
those redshifts. 

This procedure of removing catastrophic objects changes the selection 
function of the sample, which in turn changes the redshift distribution.
When culling a catalog using an estimated error, one should carefully 
consider the effects of the reduced sample size as well as the change
in the selection function of the objects to be analyzed.
Recently, there has been promising work showing that, for the DES mock
catalog, the accuracy of galaxy power spectrum measurement can be 
improved by culling high estimated error galaxies using
the MDE estimator (Banerji et al., in prep).
The study finds that the improvement in the photo-z scatter outweighs
the reduced statistics of the resulting smaller sample of low photo-z
error galaxies.

\section{Conclusions}\label{section:conclusion}

In this paper, we have introduced a new approach to estimating photometric 
redshift errors using a spectroscopic training set.
We presented two implementations of the training-set approach, Kd-tree
and Nearest Neighbor Error (NNE), and found 
that NNE is the best error estimator when a representative training set 
is available. 
Compared to the $\chi^{2}$ error and the MDE estimators, 
training-set based error estimators are less sensitive to systematic errors in 
magnitude error estimates. They incorporate both the bias and scatter of the
photo-z's, important features given the often substantial biases in photo-z estimates.
Comparison of NNE and Kd-tree errors with error estimators from the literature 
shows that these training-set error estimators are in general more accurate and better
behaved (in the sense that the error residual distribution is closer to a Gaussian).

Since a fully representative spectroscopic training set is not always 
available, we explored the impact on these error estimates of non-representative 
training sets. We found that this does not substantially 
degrade the accuracy of the training-set error estimates.
In fact, we showed that even for training sets with very different  
magnitude and redshift distributions from the photometric sample, the 
training-set error estimates remain accurate at the 10\% level. 

Finally, we demonstrated that one can cull galaxies with large estimated 
errors from a sample and thereby significantly 
improve the overall scatter and bias of the photo-z
estimates. 
Because the training-set error estimators are more accurate than other  
error estimators, and because the photo-z error residuals
are nearly Gaussian distributed for these methods, culling objects using 
NNE or Kd-tree results in greater performance improvement than culling with other 
error estimators.
 
\section{Acknowledgments}

We would like to thank Erin Sheldon for insightful and useful
discussions, as well as Dinoj Surendran and Mark SubbaRao for introducing
the authors to a fast method of nearest neighbor search using Cover-Trees.

This work was supported in part by the Kavli Institute for Cosmological
 Physics at the University of Chicago through grants NSF PHY-0114422 and
 NSF PHY-0551142 and an endowment from the Kavli Foundation and its 
founder Fred Kavli.
HO was additionally supported by the NSF grants AST-0239759, AST-0507666, 
and AST-0708154 at the University of Chicago.
ML was additionally supported by the Department of Energy grant to the 
University of Chicago and Fermilab.
Support for CC is made available through the contract between DOE and Fermi 
Research Alliance, LLC, contract number DE-AC02-07CH11359. 

Funding for the SDSS and SDSS-II has been provided by the Alfred P. 
Sloan Foundation, the Participating Institutions, the National Science 
Foundation, the U.S. Department of Energy, the National Aeronautics and
 Space Administration, the Japanese Monbukagakusho, the Max Planck 
Society, and the Higher Education Funding Council for England. The 
SDSS Web Site is {\tt http://www.sdss.org/}.

The SDSS is managed by the Astrophysical Research Consortium for the
 Participating Institutions. The Participating Institutions are the
 American Museum of Natural History, Astrophysical Institute Potsdam,
 University of Basel, University of Cambridge, Case Western Reserve
 University, University of Chicago, Drexel University, Fermilab, the
 Institute for Advanced Study, the Japan Participation Group, Johns
 Hopkins University, the Joint Institute for Nuclear Astrophysics,
 the Kavli Institute for Particle Astrophysics and Cosmology, the 
Korean Scientist Group, the Chinese Academy of Sciences (LAMOST), 
Los Alamos National Laboratory, the Max-Planck-Institute for Astronomy 
(MPIA), the Max-Planck-Institute for Astrophysics (MPA), New Mexico 
State University, Ohio State University, University of Pittsburgh, 
University of Portsmouth, Princeton University, the United States 
Naval Observatory, and the University of Washington. 

\bibliographystyle{apj}
\bibliography{ms}

\end{document}